\documentclass[sigplan,nonacm=true]{acmart}
\usepackage{listings}
\usepackage{xcolor}
\usepackage{microtype}
\usepackage{graphicx}
\usepackage{subcaption}
\usepackage{booktabs, tabularx} 

\usepackage{amssymb}
\usepackage{algorithmic}
\usepackage[ruled,vlined,linesnumbered]{algorithm2e}
\usepackage{float} 
\usepackage[mathlines, switch]{lineno}
\usepackage{listings}
\usepackage{textcomp}
\usepackage{xcolor}
\usepackage{algorithm2e}
\usepackage{enumitem}
\usepackage{amssymb}
\usepackage[table]{xcolor}
\usepackage{tikz}

\usepackage{hyperref}

\definecolor{codebg}{RGB}{248,248,248}
\definecolor{codeframe}{RGB}{220,220,220}
\definecolor{gitcolor}{RGB}{180,70,45}
\definecolor{dockercolor}{RGB}{40,120,180}
\definecolor{graphmendcolor}{RGB}{115,75,170}

\lstdefinestyle{artifactbash}{
    basicstyle=\ttfamily\scriptsize,
    backgroundcolor=\color{codebg},
    frame=single,
    rulecolor=\color{codeframe},
    framesep=5pt,
    breaklines=true,
    columns=fullflexible,
    keepspaces=true,
    showstringspaces=false,
    xleftmargin=2pt,
    xrightmargin=2pt,
    aboveskip=4pt,
    belowskip=4pt,
    literate=
      {git}{{{\color{gitcolor}\bfseries git}}}3
      {docker}{{{\color{dockercolor}\bfseries docker}}}6
      {graphmend}{{{\color{graphmendcolor}\bfseries graphmend}}}9
}

\usepackage{listings}
\usepackage{xcolor}

\definecolor{codebg}{RGB}{255,253,240}
\definecolor{codeframe}{RGB}{200,200,200}
\definecolor{keyword}{RGB}{0,0,180}
\definecolor{string}{RGB}{163,21,21}
\definecolor{comment}{RGB}{0,128,0}
\definecolor{identifier}{RGB}{0,0,0}
\definecolor{emphcolor}{RGB}{255,140,0} 

\lstdefinestyle{mypython}{
    language=Python,
    backgroundcolor=\color{codebg},
    basicstyle=\ttfamily\scriptsize, 
    keywordstyle=\color{keyword}\bfseries,
    stringstyle=\color{string},
    commentstyle=\itshape\color{comment},
    identifierstyle=\color{identifier},
    numbers=left,
    numberstyle=\tiny\color{gray},
    stepnumber=1,
    numbersep=8pt,
    frame=single,
    rulecolor=\color{codeframe},
    tabsize=4,
    showstringspaces=false,
    breaklines=true,
    breakatwhitespace=false,
    keepspaces=true,
    columns=fullflexible,
    linewidth=\linewidth,
    xleftmargin=0.5cm,
    xrightmargin=0.1cm,
    breakindent=0pt,
    prebreak=\mbox{\tiny$\hookleftarrow$},
    postbreak=\mbox{\textcolor{gray}{$\hookrightarrow$}\space},
    captionpos=b,
    emph={Tensor, torch, nn, autograd},
    emphstyle=\color{emphcolor}\bfseries,
    breaklines=true,
    postbreak={}
}

\lstnewenvironment{python}[1][]{
  \lstset{style=mypython,#1}
}{}

\newcommand{\pyinline}[1]{%
  \lstinline[
    style=mypython,
    basicstyle=\ttfamily\footnotesize
  ]!#1!%
}

\usepackage{listings}
\usepackage{xcolor}

\definecolor{outbg}{RGB}{245,248,255}
\definecolor{outframe}{RGB}{200,200,200}
\definecolor{outtext}{RGB}{40,40,40}

\lstdefinestyle{myoutput}{
    backgroundcolor=\color{outbg},
    basicstyle=\ttfamily\footnotesize\color{outtext}, 
    numbers=none,
    frame=single,
    rulecolor=\color{outframe},
    showstringspaces=false,
    breaklines=true,
    breakatwhitespace=false,
    captionpos=b,
    columns=flexible,
    keepspaces=true,
    xleftmargin=0.5cm
}

\lstnewenvironment{out}[1][]{
  \lstset{style=myoutput,#1}
}{}

\AtBeginDocument{%
  }

\setcopyright{none}
\copyrightyear{2026}
\acmYear{2026}
\acmConference[CGO 2027]{International Symposium on Code Generation and Optimization}{}{Salt Lake City, Utah}

\begin{document}

\title{GraphMend: Code Transformations for Fixing Graph Breaks in PyTorch 2}

\author{Savini Kashmira}
\affiliation{%
      \institution{University of Michigan}
            \country{USA}}
\email{savinik@umich.edu}

\author{Jayanaka L. Dantanarayana}
\affiliation{%
      \institution{University of Michigan}
      \country{USA}}
\email{jayanaka@umich.edu}

\author{Thamirawaran Sathiyalogeswaran}
\affiliation{%
      \institution{Jaseci Labs}
      \country{USA}}
\email{thami@jaseci.org}

\author{Krisztian Flautner}
\affiliation{%
      \institution{University of Michigan}
      \country{USA}}
\email{manowar@umich.edu}

\author{Lingjia Tang}
\affiliation{%
      \institution{University of Michigan}
      \country{USA}}
\email{lingjia@umich.edu}

\author{Jason Mars}
\affiliation{%
      \institution{University of Michigan}
      \country{USA}}
\email{profmars@umich.edu}

\renewcommand{\shortauthors}{Savini Kashmira et al.}

\begin{abstract}

This paper presents \textsc{GraphMend}, a compiler technique that automatically fixes FX graph breaks in PyTorch 2 programs. Although PyTorch 2 introduced TorchDynamo and TorchInductor to enable just-in-time graph compilation, certain code patterns still cause graph breaks that force execution to fall back to Python eager mode, introducing costly CPU-GPU synchronization and reducing optimization opportunities. Our investigation of 195 Hugging Face models reveals that 13.8\% of models exhibit graph breaks. \textsc{GraphMend} automatically eliminates fixable breaks through source-level program analysis and transformations. It analyzes AST-level program structure to identify graph-break patterns and applies transformations only when their semantic preservation can be statically established. These transformations enable PyTorch to capture larger, uninterrupted FX graphs without manual refactoring by developers. We evaluate \textsc{GraphMend} on all 27 models found to exhibit graph breaks in our investigation. \textsc{GraphMend} eliminates 107 of 147 graph breaks (73\%), fully fixing all breaks in 21 models. In our experiments on NVIDIA GPUs, \textsc{GraphMend} achieves up to 26$\times$ cold-start forward pass speedup (5$\times$ on average) and up to 1.39$\times$ steady-state forward pass speedup. These results demonstrate that semantics-aware source-level analysis and transformation are effective complements to PyTorch's dynamic JIT compilation pipeline, substantially improving both usability and performance.

\end{abstract}




\maketitle

\makeatletter
\begingroup
  \let\@makefnmark\relax
  \let\@thefnmark\relax
  \let\@makefntext\noindent
  \footnotetextcopyrightpermission{%
    \parbox[t]{\dimexpr0.5\textwidth-0.5\columnsep\relax}{\footnotesize
      Accepted for publication at the International Symposium on Code
      Generation and Optimization (CGO 2027), Salt Lake City, Utah.}}%
\endgroup
\makeatother

\section{Introduction}

PyTorch has emerged as one of the most widely adopted deep learning frameworks   in both academia and industry, primarily due to its ease of use and flexibility~\cite{pytorch}. However, its general-purpose design incurs efficiency overheads, leaving PyTorch workloads less optimized than handcrafted GPU implementations. 
To bridge this gap, PyTorch 2~\cite{pyotrch2} introduced a Python Just-in-Time (JIT) compilation pipeline that translates model operations into optimized GPU code, achieving performance close to handcrafted implementations. 

One of the central intermediate representations (IRs) in this pipeline is the FX graph~\cite{torch.fx}. TorchDynamo intercepts and symbolically evaluates Python bytecode to create an FX graph. The FX graph is then passed to a backend compiler, by default, TorchInductor~\cite{pyotrch2}, which applies a series of graph-level and kernel-level optimizations before lowering the computation to efficient kernels for GPU execution. When tracing the FX graph, TorchDynamo often struggles with unsupported Python operations, which it handles by falling back to eager mode. These unsupported operations include data-dependent control flow such as conditionals that depend on runtime values, certain Python side effects like I/O operations, and runtime validation guards that perform data-dependent assertions. In such cases, it inserts graph breaks, resulting in multiple disjoint FX graphs instead of a single unified one~\cite{GRAPE, pyotrch2}.


%

Graph breaks have been identified as a key challenge in realizing the full performance potential of the PyTorch~2~\cite{pytorch_compiler_faq, pytorch_compile_tutorial, rand2025pytorch_compile, hoyath2025pytorch_dynamo}. Although the resulting disjoint FX graphs are still optimized by the backend, the regions between them execute in PyTorch eager mode~\cite{pytorch}. This repeatedly returns control to the Python interpreter, introducing overhead from Python dispatch and breaking the continuous compiled execution stream. In many cases, these transitions also force CPU-GPU synchronization, especially when the break is caused by data-dependent control flow or other runtime checks that require CPU-side evaluation. Moreover, each FX graph is compiled independently, preventing cross-graph fusion and global optimization~\cite{agrawal2019tensorfloweagermultistagepythonembedded, pyotrch2}. During the cold run, each disjoint FX graph must also be separately traced and optimized, multiplying compilation overhead proportionally to the number of breaks. When CUDA Graphs are employed, which is common in inference serving, each subgraph must additionally be recorded as a separate CUDA graph. These costs also limit the benefits of modern serving frameworks such as vLLM and SGLang, where graph breaks reduce the effectiveness of \pyinline{torch.compile}-based graph capture and CUDA Graph replay for inference acceleration~\cite{vllm-debug-compile, vllm_piecewise_compile, sglang_server_args}.

To further understand the causes and remedies of graph breaks, we examine 195 models from Hugging Face~\cite{huggingface2025}, taking the top 100 trending models, the top 30 by downloads, and 65 randomly sampled models spanning diverse architectures and task categories. We find that 27 models (13.8\%) exhibit graph breaks, collectively containing 147 individual break instances. These include some of the most widely used models on the platform: Google's T5-small with over 2.8M monthly downloads, OpenAI's Whisper-small with over 2M, Facebook's Bart-large-cnn with 2.72M, and Microsoft's Phi-4-mini-instruct with over 584k (as of March 2026). This demonstrates that graph breaks are not merely a consequence of developer inexperience or poor coding practices, but a structural challenge common even in production models maintained by leading AI organizations. 

We taxonomize the common causes of graph breaks in the affected models and find that many breaks arise from high-level source-code patterns, such as data-dependent control flow and unsupported Python constructs like logger calls and validation guards. Such structure is often lost once the code is lowered to bytecode. Since PyTorch's current compilation pipeline begins at the bytecode level, tracing instruction by instruction, it lacks the semantic context needed to resolve these breaks. Since these source-level patterns account for 73\% of the graph breaks found in our study, our key insight is that \textbf{graph breaks should be addressed at the source level, before bytecode lowering, where the relevant program structure is still available, making it feasible to eliminate graph breaks by transforming Python code into forms that use PyTorch-supported operations, allowing the compiler to capture more of the program as a single, continuous FX graph.}

Although PyTorch documents source-level fixes such as \pyinline{torch.where} and \pyinline{torch.cond}, applying them requires developers to perform analysis to establish semantic equivalence and select a safe transformation.
As shown in Figure~\ref{fig:phi-4-breaks}, fixing a single break in Phi-4-mini-instruct may require hoisting branches, reconciling different function calls, and refactoring side effects, which in turn requires reasoning about control flow, data dependencies, side effects, exceptions, and state mutations. This is compiler-level
reasoning that should not be the burden of model developers.

To automate the analysis and source-level fixes, we introduce \textsc{\textbf{GraphMend}}, a compiler technique that eliminates graph breaks through high-level program analysis and transformation. \textsc{GraphMend} extends the standard PyTorch 2 compilation pipeline with an AST-level transformation phase that restructures Python code before bytecode generation, improving traceability by eliminating graph breaks. Each transformation is guarded by a conservative legality analysis over the AST, control-flow graph, and symbol table, and is applied only when its semantic-preservation conditions can be statically established; otherwise, the code is left unchanged. Subject to this analysis, \textsc{GraphMend} applies three transformations: (1) Predicated Data-Dependent Control Flow, which rewrites data-dependent branches into \pyinline{torch.where} expressions that keep both paths inside the FX graph; (2) Graph-Epilogue Deferred Side Effects, which buffers \pyinline{print} and \pyinline{logger} calls and flushes them after graph execution; and (3) Predicated Trap Lowering, which replaces \pyinline{if not condition: raise} patterns with \pyinline{torch._assert_async}, a graph-native assertion that preserves runtime safety checks without breaking tracing.



We implement \textsc{GraphMend} within the Jaseci framework~\cite{JAC_CAL, jac_OSP, jaseci2025}, a front-end compiler that accepts standard Python programs. Since Jaseci compiles Python source code into native Python bytecode and executes it on the standard Python interpreter, all PyTorch models run directly without any code changes. Jaseci constructs an abstract syntax tree (AST), a control flow graph (CFG), and a symbol table, merging them into a unified intermediate representation (UniiR)~\cite{mtp}, which enables rich program analysis and transformation at the source level. We leverage this capability and extend the Jaseci compiler with our own passes that detect and restructure code patterns prone to graph breaks before bytecode generation.

From our survey of 195 Hugging Face~\cite{huggingface2025} models, we evaluate \textsc{GraphMend} on all 27 that exhibit graph breaks. \textsc{GraphMend} eliminates all fixable breaks in 21 models, partially fixes 3 others, and leaves only 3 models unfixed whose breaks stem from fundamentally untraceable operations such as dynamic shape operators. Across all 27 models, \textsc{GraphMend} removes 107 of 147 graph breaks (73\%). On NVIDIA RTX 3090, A40 and H100 GPUs, it achieves up to 26$\times$ cold-start forward pass speedup (5$\times$ on average) and up to 1.39$\times$ steady-state forward pass speedup.


Our contributions are: (1) A large-scale empirical study of graph break prevalence across 195 Hugging Face models, quantifying the scope of the problem in production models. (2) A source-level analysis that identifies fixable graph-break patterns and uses the AST, CFG, and symbol table to establish conservative legality conditions for semantics-preserving transformations. (3) The \textsc{GraphMend} compiler technique, which applies three automated AST-level transformations to eliminate graph breaks before bytecode generation. (4) A comprehensive evaluation across 27 models on NVIDIA GPUs, with detailed profiler-based analysis of cold-start compilation overhead, CPU-GPU synchronization behavior, and kernel fusion opportunities exposed by graph break elimination.

\vspace{-0.1cm}

\section{Motivation}
\label{sec:motivation_example}
\begin{figure*}[t]
    \centering
    \begin{subfigure}[t]{0.9\textwidth}
        \includegraphics[width=\linewidth]{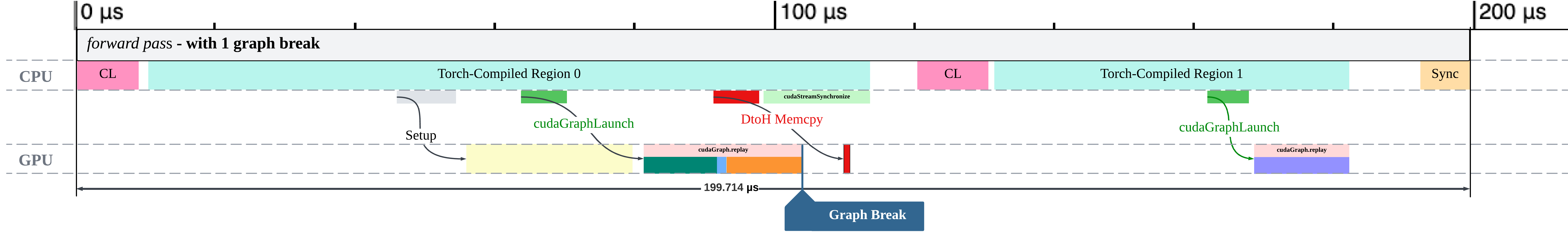}
        \caption{}
        \label{fig:trace_with_breaks}
        \end{subfigure}
           \vspace{-0.1cm}
    \begin{subfigure}[t]{0.9\textwidth}
        \includegraphics[width=\linewidth]{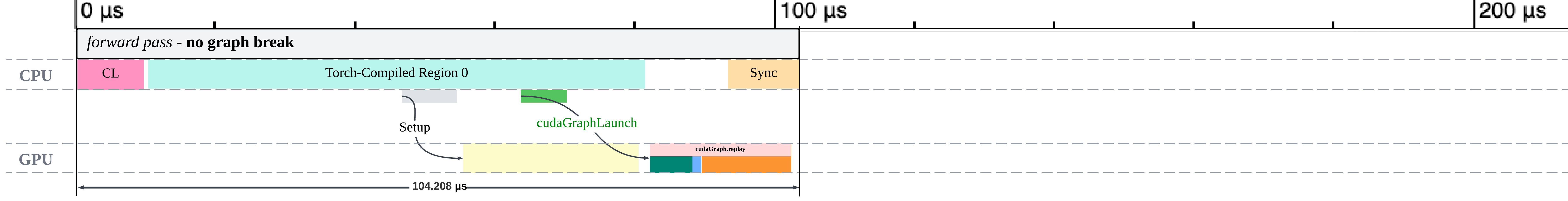}
        \caption{}
        \label{fig:trace_fixed}
    \end{subfigure}
    \vspace{-0.1cm}
    \caption{Profiled traces of forward pass execution across CPU and GPU. (a) Forward pass execution trace of code with graph breaks in Figure~\ref{fig:dc_example}(a). (b) Forward pass execution trace of equivalent code with graph breaks fixed in Figure~\ref{fig:dc_example}(b).}
    \label{fig:profiler_data}
\end{figure*}

\begin{figure}[t]
    \centering
\begin{python}
@torch.compile()
def f(x, y):
    x_1 = x*2
    y_1 = y*2
    if x.sum() > 0: # <-- graph-break here
        z = x_1 + y_1
    else:
        z = x_1 * y_1
    return torch.relu(z)
\end{python}
    {\small \textbf{(a)} Original: data-dependent branch causes a graph break}

    \vspace{0.5em}

\begin{python}
@torch.compile()
def f(x, y):
    x_1 = x*2
    y_1 = y*2
    cond = x.sum() > 0
    z_add = x_1 + y_1
    z_mul = x_1 * y_1
    z = torch.where(cond, z_add, z_mul)
    return torch.relu(z)
\end{python}
    {\small \textbf{(b)} Transformed: \pyinline{torch.where} keeps a single FX graph}
    \caption{A PyTorch forward pass with data-dependent control flow (a), and the rewritten version (b) where \pyinline{torch.where} eliminates the graph break.}
    \label{fig:dc_example}
    \vspace{-0.5cm}
\end{figure}

\paragraph{\textbf{FX graph breaks.}}

When TorchDynamo tries to symbolically evaluate the function in Figure~\ref{fig:dc_example}(a) to capture an FX graph, it cannot capture the conditional branch because the outcome depends on the runtime value of the tensor expression \pyinline{x.sum()}. Therefore, it inserts a graph break~\cite{pyotrch2, pytorch_compiler_troubleshooting}.

\vspace{-0.15cm}

\begin{figure*}[ht]
    \centering
    \begin{subfigure}[t]{0.9\linewidth}
        \centering
        \includegraphics[width=\linewidth]{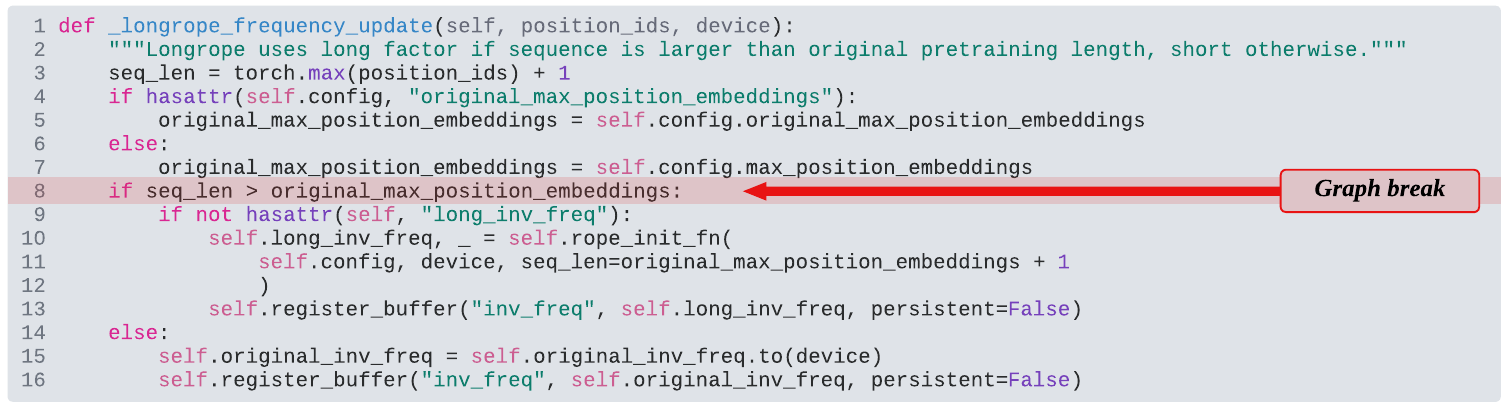}
        \caption{Existing graph-break in Phi-4-mini instruct model built by Microsoft.}
        \label{fig:phi-4_break}
    \end{subfigure}
    \begin{subfigure}[t]{0.9\linewidth}
        \centering
        \includegraphics[width=\linewidth]{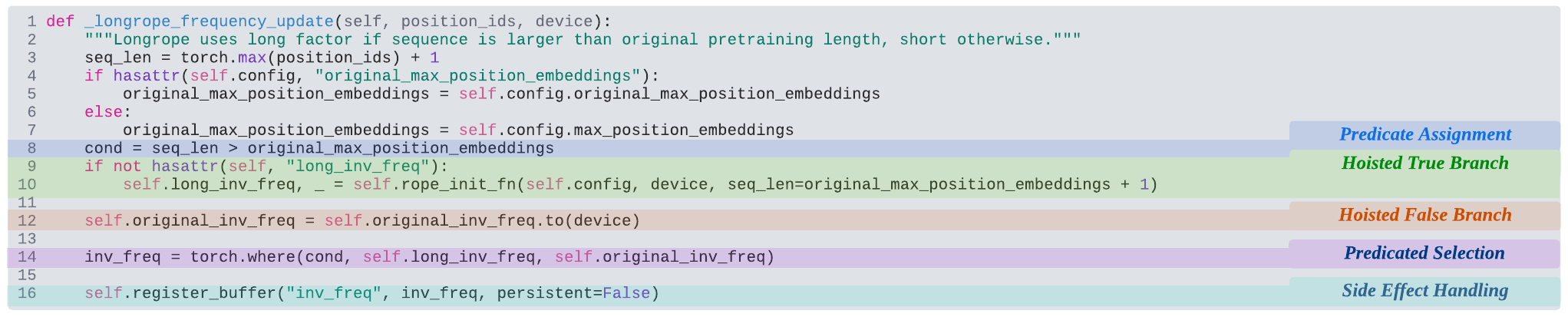}
        \caption{Version of Phi-4-mini instruct model with the graph break fixed, without affecting program correctness.}
        \label{fig:phi-4_fix}
    \end{subfigure}
    \caption{(a) Existing graph-breaks in a real model and (b) how it can be fixed to produce a single contiguous FX graph. }
    \label{fig:phi-4-breaks}
\end{figure*}

\vspace{-0.1cm}

\paragraph{\textbf{Impact of graph breaks.}}
\label{graph_break_impact}

When graph breaks occur, the function is divided into multiple FX graphs, each compiled separately into GPU kernels~\cite{pyotrch2, ghosh2025pygraphrobustcompilersupport}. During execution, control returns to the Python interpreter between compiled regions to execute the unsupported code in eager mode, introducing CPU--GPU synchronization overhead and reducing GPU utilization~\cite{pyotrch2, pytorch_compiler_troubleshooting}.

Figure~\ref{fig:trace_with_breaks} illustrates this effect for the function in Figure~\ref{fig:dc_example}. The trace shows two separate \pyinline{torch.compile} regions divided by a device-to-host (D2H) memory transfer, where the GPU remains idle until the next CUDA graph launch. This fragmentation increases CPU scheduling overhead and prevents optimizations such as kernel fusion. When the break is removed and captured as a single FX graph, the execution trace in Figure~\ref{fig:trace_fixed} shows one continuous CUDA region with full GPU utilization and no intermediate D2H transfers.

\paragraph{\textbf{Why identifying and fixing graph breaks is hard.}}
 
To illustrate the difficulty of manually fixing graph breaks, we examine Phi-4-mini-instruct~\cite{microsoft_phi4_mini_instruct}, a widely used model from Microsoft. We profiled it using PyTorch~2 and identified 5 graph breaks caused by data-dependent control flow. Figure~\ref{fig:phi-4-breaks} shows such a case.
 
Identifying which code causes the break is not straightforward. The function in Figure~\ref{fig:phi-4_break} contains 3 \pyinline{if-else} blocks at lines~4, 8, and~9. The \pyinline{if} at line~4 checks \pyinline{hasattr(self.config, ...)}, which is a static attribute check that Dynamo can handle without a break. The \pyinline{if} at line~9 is also a \pyinline{hasattr} check, so no break there either. Only line~8 causes a graph break because it compares \pyinline{seq\_len} against a threshold. But to understand why, the developer must trace back to line~3, where \pyinline{seq\_len} is assigned from \pyinline{torch.max(position\_ids)}, a tensor-dependent operation. This is what makes the predicate at line~8 data-dependent and untraceable by Dynamo.
 
Fixing the break is equally difficult. A simple \pyinline{torch.where} replacement does not work here because the true and false branches call entirely different functions: the true branch calls \pyinline{rope\_init\_fn} (line~10) while the false branch calls \pyinline{.to(device)} (line~15). As shown in Figure~\ref{fig:phi-4_fix}, the fix requires extracting the condition into a predicate variable (line~8), hoisting both branch computations out of the conditional (lines~9-12), selecting between their results using \pyinline{torch.where} (line~14), and moving the \pyinline{register\_buffer} side effect to execute once with the predicated result (line~16). Each of these steps requires reasoning about data dependencies, control flow, and side effects.
 
After applying this fix, all 5 graph breaks were eliminated and profiling showed up to 4.3$\times$ cold-start forward pass speedup and 1.14$\times$ steady-state forward pass speedup (further discussed in \S~\ref{sec:eval}). This example shows that while the fixes are effective, they require compiler-level reasoning that is impractical to expect from model developers, especially when the same patterns appear across many models.
\section{Understanding and Eliminating Common Graph Breaks}
\label{sec:graph_breaks}



\begin{figure}[t]
    \centering
\begin{python}
@torch.compile
def fn(x):
    x = torch.relu(x)
    print("tensor:", x)  # <-- graph-break here
    return torch.sin(x)
\end{python}
    {\small \textbf{(a)} Original: \pyinline{print} causes a graph break}

    \vspace{0.5em}

\begin{python}
@torch.compile
def fn(x):
    x = torch.relu(x)
    to_print = "tensor:", x  # <-- assign variable to print
    y = torch.sin(x)
    print(to_print)  # <-- print at the end of tracing
    return y
\end{python}
    {\small \textbf{(b)} Transformed: \pyinline{print} deferred past the traced region}
    \caption{A side-effecting \pyinline{print} call that causes a graph break (a), and the transformed version (b) where the value is buffered during tracing and flushed after the graph executes.}
    \label{fig:sideeffect_example}
    \vspace{-0.3cm}
\end{figure}


\begin{figure}[t]
    \centering
\begin{python}
@torch.compile
def fn(attention_mask, expected):
    if not torch.equal(attention_mask, expected):
        raise ValueError(  # <-- graph-break here
            "unsupported attention mask")
    return process(attention_mask)
\end{python}
    {\small \textbf{(a)} Original: \pyinline{torch.equal} guard causes a graph break}

    \vspace{0.5em}

\begin{python}
@torch.compile
def fn(attention_mask, expected):
    _meta_ok = (attention_mask.shape == expected.shape
                and attention_mask.dtype == expected.dtype)
    _cond = (
        (attention_mask == expected).all()
        if _meta_ok
        else torch.tensor(False, device=attention_mask.device)
    )
    torch._assert_async(_cond, 'unsupported attention mask')
    return process(attention_mask)
\end{python}
    {\small \textbf{(b)} Transformed: \pyinline{torch.\_assert\_async} keeps the check in the graph}
    \caption{A validation guard using \pyinline{torch.equal} whose data-dependent bool result causes a graph break (a), and the transformed version (b).}
    \label{fig:guard_example}
    \vspace{-0.5cm}
\end{figure}

\begin{figure*}
    \centering
    \includegraphics[width=0.97\linewidth]{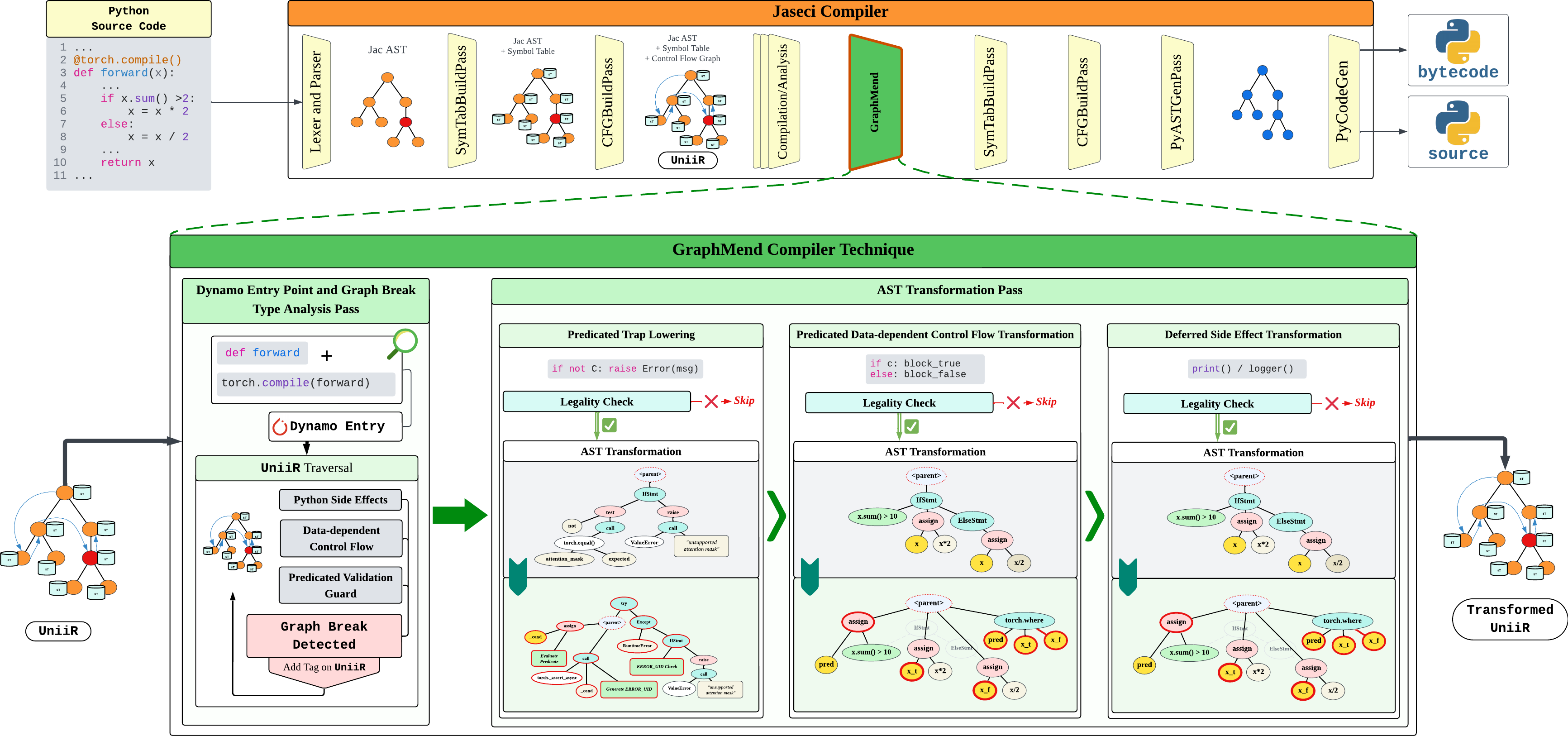}
    \caption{GraphMend compiler integration in the Jaseci pipeline. The pipeline (top) lowers Python source code into a unified intermediate representation (UniiR) with AST, symbol table, and CFG. GraphMend (bottom) analyzes entry points, detects graph breaks, and applies AST transformations to eliminate breaks before execution.}
    \label{fig:compiler}
\end{figure*}

In this work, we focus on three common causes of graph breaks, outlining code transformations to eliminate them and explaining why bytecode-level fixes are challenging without higher-level IR context.
\vspace{-0.2cm}

\paragraph{\textbf{Data-dependent Control Flow.}}

In PyTorch~2, symbolic tracing builds FX graphs using \pyinline{FakeTensor} objects, assuming control flow decisions can be inferred from static information such as tensor shapes or constants. When a condition depends on actual tensor data (for example, \pyinline{if x.sum() > 0:} in Figure~\ref{fig:dc_example}(a)), the compiler cannot safely evaluate it, so PyTorch inserts a graph break. We can fix them by expressing data-dependent branches using tensor operations such as \pyinline{torch.where} or \pyinline{torch.cond}, which are supported on GPU. Figure~\ref{fig:dc_example} shows the rewritten version of the same function, where the use of \pyinline{torch.where} eliminates the graph break.



\paragraph{\textbf{Python Side-effects.}}
Side-effecting operations such as \pyinline{print} or \pyinline{logging} calls also cause graph breaks because they interact with the Python runtime and cannot be represented in a functional graph. Deferring these effects until after the graph executes can remove the break while preserving program behavior. In Figure~\ref{fig:sideeffect_example}, we show how moving a \pyinline{print} statement to the end of the function allows the entire function to remain in one FX graph.

PyTorch provides a built-in mechanism for this, \pyinline{torch._dynamo.config.reorderable_logging_functions}, which defers registered logging calls until after the compiled region~\cite{pytorch_common_graph_breaks}. However, it did not eliminate any of the logger-call breaks in our benchmark suite. Its eligibility check accepts only plain and builtin function types, rejecting bound logger methods~\cite{pytorch_issue_132635}. Eliminating these breaks while preserving the logging output therefore requires deferring the calls at the source level, where the callable can be resolved.

\paragraph{\textbf{Predicated Validation Guards.}}
A common pattern in real-world models is to validate tensor properties before proceeding with computation, for example checking that an attention mask matches the expected shape using \pyinline{torch.equal} (Figure~\ref{fig:guard_example}(a)). Unlike data-dependent control flow, validation guards raise exceptions on failure and produce no tensor output, making \pyinline{torch.where} and \pyinline{torch.cond} unsuitable.
 
We fix these by replacing the Python guard with \pyinline{torch._assert_async}, a graph-native assertion that raises an error at runtime without breaking tracing. As shown in Figure~\ref{fig:guard_example}(b), the transformation has two steps. First, a Python-level check verifies matching shapes and dtypes. This evaluates to a Python bool that Dynamo guards on statically, introducing no graph break. This step is necessary because element-wise comparison \pyinline{(a == b).all()} uses broadcasting and would incorrectly return \pyinline{True} for tensors with different but broadcastable shapes, whereas \pyinline{torch.equal} would return \pyinline{False}. Second, if shapes and dtypes match, \pyinline{torch.equal(a, b)} is rewritten as \pyinline{(a == b).all()}, producing a tensor that is passed to \pyinline{torch._assert_async}.


\paragraph{\textbf{Breaks outside the scope of source-level transformation.}}

Not all graph breaks can be eliminated by rewriting source code. Two
categories among the breaks in Table~\ref{tab:benchmark-merged} resist
semantics-preserving source-level transformation. The first is
\pyinline{tensor.item()} calls, which extract a Python scalar from a
tensor; the resulting value flows into the Python interpreter, so the
computation that follows cannot stay inside the FX graph regardless of
how the call is expressed. The second is dynamic shape and
data-dependent operators such as \pyinline{torch.nonzero} or boolean
mask indexing, whose output shapes depend on tensor values; since FX
graph capture requires shapes to be known symbolically at trace time,
no rewrite of the source can remove this data dependence. These breaks
require runtime values that have no tensor-level equivalent, and we
leave them outside \textsc{GraphMend}'s scope.

\vspace{-0.2cm}
\paragraph{\textbf{Why automating these transformations at the bytecode level is challenging}}

Automating these transformations must occur before PyTorch's bytecode analysis, but performing safe transformations at the bytecode level is challenging since much of the high-level semantic information is lost during compilation.

When high-level constructs such as \pyinline{if/else} or \pyinline{while} loops are lowered to bytecode, they become low-level jump instructions like \texttt{POP\_JUMP\_IF\_FALSE} and \texttt{JUMP\_BACKWARD}. Since these opcodes are reused across different control-flow patterns, distinguishing loops from conditionals and identifying their bodies is non-trivial and often requires partial reverse engineering of the bytecode. Ensuring correctness during transformation further complicates this process.

Moreover, performing data-flow analysis at the bytecode level is difficult since Python uses a stack-based rather than a register-based model. While variable operations themselves are explicit, intermediate expression values are pushed and popped from the operand stack without explicit naming, obscuring the flow of computed values without symbolic stack simulation. This simulation adds complexity and overhead to the analysis, whereas analyses on higher-level AST representations with explicit variable bindings and control flow are far simpler.

Handling side effects at the bytecode level is also challenging. While delaying operations like built-in \pyinline{print} or \pyinline{logger} calls may be safe, Python's dynamic nature allows users to redefine these functions. Determining whether a call refers to the original built-in or a user-defined override requires tracking symbol bindings, which is only possible with higher-level semantic context.

A higher-level IR that retains program structure and data dependencies is therefore better suited for automatically applying these transformations.

\section{\textsc{GraphMend} Compiler Technique}
\label{sec:graphmend}

In this section, we introduce our compiler technique \textsc{GraphMend}, which automates these code transformations using AST level analyses and transformations, to eliminate graph breaks, while maintaining semantic equivalency. The implementation of this technique is shown in Figure~\ref{fig:compiler}, which is elaborated in the rest of this section.

\subsection{Jaseci Compilation Pipeline}

To implement our compiler technique, we build on the Jaseci framework~\cite{JAC_CAL,jac_OSP,mtp}. Jaseci accepts standard Python programs and produces native Python bytecode, \textbf{requiring no changes to existing model code}. It is distributed as a PyPI package, allowing users to install and run their models with a single command. As shown in the top half of Figure~\ref{fig:compiler}, Jaseci parses the source into an AST and progressively lowers it into UniiR, a unified intermediate representation that combines the AST, CFG, and symbol table into a single structure, enabling rich syntactic, semantic, and control-flow analysis.

\textbf{Our \textsc{GraphMend} compiler technique} consists of two passes built into the Jaseci compilation pipeline, illustrated in the bottom half of Figure~\ref{fig:compiler}: an analysis pass that identifies TorchDynamo entry points and classifies fixable graph breaks, followed by an AST transformation pass that performs legality checks and applies transformations in a fixed order.

\subsection{Dynamo Entry-Point and Break Analysis Pass}
\label{sec:graphmend-analysis}

The analysis pass first locates functions or \pyinline{nn.Module}
objects wrapped or decorated with \pyinline{torch.compile} and tags
them as Dynamo entry points; if none exist, \textsc{GraphMend} hands
off to the regular compilation pipeline. Within each entry point, it
then tags every graph-break candidate with its category, following
the source-level patterns of Section~3. A predicate is data-dependent
when its truth value depends on runtime tensor contents: either it
directly applies a Torch operation whose result depends on tensor
values, such as \pyinline{.sum()}, or use-def analysis over the
symbol table and CFG~\cite{aho2006compilers} traces a name in it to
such an operation. Function parameters are not assumed
data-dependent, so static checks such as \pyinline{hasattr} remain
untouched.

A conditional with a data-dependent predicate is tagged as a
validation guard when its body is a single \pyinline{raise}
statement, and as data-dependent control flow otherwise;
\pyinline{print} and supported logger calls are tagged as side-effect
breaks.

These tags select the rules of Table~\ref{tab:graphmend-rules}. For the rules, we write $\mathit{tensorize}(C)$ for a tensor-valued Boolean with the same truth value as the Python condition $C$; when no conversion is known, $\mathit{tensorize}(C)$ is undefined.

\begin{table*}[t]
\centering
\caption{\textsc{GraphMend} transformation rules. A rule is applied
only when all listed legality conditions hold; otherwise, the original
region is preserved.}
\label{tab:graphmend-rules}
\vspace{-0.3cm}

\fontsize{8pt}{9pt}\selectfont
\setlength{\tabcolsep}{4pt}
\renewcommand{\arraystretch}{1.18}

\begin{tabular}{@{}
p{0.10\textwidth}
p{0.42\textwidth}
p{0.42\textwidth}
@{}}
\toprule
\textbf{Rule}
&
\textbf{Rewrite}
&
\textbf{Legality conditions}
\\
\midrule

\textsc{[Trap]}
&
\texttt{if not C: raise E(msg)}
\newline
\hspace*{1em}$\Rightarrow$
\texttt{torch.\_assert\_async(}%
$\mathit{tensorize}(C)$%
\texttt{, uid + msg)}
&
The guard body contains only the \texttt{raise}, and
$\mathit{tensorize}(C)$ is defined.
\\
\addlinespace

\textsc{[Where]}
&
\texttt{if c: }$S_t$\texttt{; }$K[e_t]$
\newline
\texttt{else: }$S_f$\texttt{; }$K[e_f]$
\newline
\hspace*{1em}$\Rightarrow$
\texttt{p = c; hoist(}$S_t$\texttt{); hoist(}$S_f$\texttt{);}
\newline
\hspace*{2.4em}$K[\texttt{torch.where}(p, e_t, e_f)]$
&
The selected values $e_t$, $e_f$ are tensor expressions with matching
shape and dtype, and neither is \texttt{None}. Every function call in
the region resolves through UniiR to inspectable code that the
analysis validates as free of observable side effects, exceptions,
and graph-break inducers.
\\
\addlinespace

\textsc{[Defer]}
&
$S_1$\texttt{; f(args); }$S_2$\texttt{; return e}
\newline
\hspace*{1em}$\Rightarrow$
$S_1$\texttt{; buf.push(f, clone(args)); }$S_2$\texttt{;}
\newline
\hspace*{2.4em}\texttt{r = e; flush(buf); return r}
&
The callee resolves to the built-in \texttt{print} function or a
supported logger and has not been rebound. Its only observable effect is console or log output.
\\

\bottomrule
\end{tabular}

\vspace{3pt}

\begin{minipage}{0.98\textwidth}
\fontsize{8.3pt}{9.6pt}\selectfont

\textbf{Notation.}
$S_t$ and $S_f$ are the \emph{setup statements} of the true and false
branches which \textsc{[Where]} executes.
$e_t$ and $e_f$ are the values produced by those branches.
$K$ is the common operation that consumes the branch value, namely an
assignment, return, or function call.
$p$ stores the branch predicate.
\texttt{uid} identifies failures introduced by \textsc{[Trap]}.
\texttt{buf} stores deferred effects, \texttt{clone} preserves argument
values at the original call site, and \texttt{flush} replays the
effects in their original order.

\end{minipage}
\vspace{-0.2cm}
\end{table*}

\subsection{AST Transformation Pass}
\label{sec:graphmend-transform}

The AST transformation pass applies the rules in
Table~\ref{tab:graphmend-rules} to the tagged graph-break regions.
Each rule fires only when all of its legality conditions can be
established from the UniiR. Otherwise, the
original region is preserved.

A setup statement is \emph{hoistable} when running it on the untaken
path changes nothing the program can observe. It must contain no
\pyinline{return}, \pyinline{break}, or \pyinline{continue}, and must
not raise on the values reaching it. Any write it performs must also
be unobservable on the untaken path: the written value is read only
at the branch join, is never read on the opposite path as verified by
an escape check over UniiR, or the write is idempotent, as with a
device move or an initialization guarded by its own existence check.
A statement may contain function calls when each callee resolves
through UniiR and its body passes the same checks.

\subsubsection{\textbf{Predicated Trap Lowering Transformation}}

Rule \textsc{[Trap]} replaces a validation guard with a graph-native
assertion. For \pyinline{torch.equal(a,b)}, \textsc{GraphMend}
first checks that the tensors have the same shape and dtype, then
converts the condition to \pyinline{(a == b).all()}. The metadata
check is required because element-wise comparison may broadcast
tensors with mismatched shapes, whereas \pyinline{torch.equal}
returns false.

\textsc{GraphMend} preserves the original exception type and message
by prefixing the generated assertion message with a unique identifier
and wrapping the compiled call in an exception handler (Table~\ref{tab:graphmend-rules}). Because
\pyinline{torch._assert_async} raises a \pyinline{RuntimeError}, the
handler recognizes the identifier and re-raises the original
exception type with the original message. Unrelated runtime errors
are propagated unchanged. The assertion remains inside the FX graph,
although its failure may be reported at the next synchronization
point rather than at the original guard.

\subsubsection{\textbf{Predicated Data-dependent Control Flow Transformation}}

Rule \textsc{[Where]} handles two-way \pyinline{if}/\pyinline{else}
conditionals whose branches
(1) assign the same variable,
(2) return compatible tensor values, or
(3) call the same function with different positional arguments (Table~\ref{tab:graphmend-rules}).

The rewrite replaces the conditional with a single
\pyinline{torch.where} selection, which fuses with the surrounding
tensor operations. After the rewrite, every statement in the region
executes on both paths, so each must be safe to run.
For a function call within a branch, this is established by resolving the callee
through UniiR and analyzing its body: the call is permitted when it
is free of observable side effects, exceptions, and graph-break
inducers, or when its only write is licensed, as with an
initialization guarded by its own existence check. A call that cannot be
resolved and validated blocks the rule. \textsc{GraphMend} checks
these conditions before rewriting; if any cannot be established, the
original conditional is preserved.

The Phi-4 example from
Figure~\ref{fig:phi-4-breaks} illustrates this process.
\textsc{GraphMend} preserves the static \pyinline{hasattr} guard,
hoists the memoized initialization and the idempotent device move,
selects the differing tensor with \pyinline{torch.where}, and emits
the shared \pyinline{register_buffer} call once.

PyTorch also provides \pyinline{torch.cond}, whose branches may call different functions but must be pure and fully 
capturable~\cite{torchcond}, so the memoized initialization and device moves in fixable branches would need hoisting regardless.
On the hoisted form, both operators remove the break, but \pyinline{torch.cond} adds CPU-GPU synchronization for branch dispatch and disables CUDA Graph capture for its region \cite{torchcondimpl} whereas \pyinline{torch.where} keeps selection on the GPU with a single launch.
\pyinline{torch.cond} may still suit heavy divergent branches; none occurred in our suite (\S~\ref{subsec:warm}).

\subsubsection{\textbf{Deferred Side Effect Transformation}}

Rule \textsc{[Defer]} applies to calls that resolve to the built-in
\pyinline{print} function or a supported logger and whose return values
are unused. These calls only produce console or log output and do not
affect program state or computed values. \textsc{GraphMend} verifies
their bindings through the symbol table. If a function has been
reassigned or shadowed, the call is left unchanged.

At the original call site, \textsc{GraphMend} records the callee and
its evaluated arguments. Tensor arguments are cloned to preserve their
values against later mutations. The recorded calls are replayed in FIFO
order after the compiled tensor computation finishes. Calls inside a
conditional are recorded only when that path executes. If the enclosing
function returns a value, \textsc{GraphMend} stores the result, replays
the deferred calls, and then returns the stored value.

\subsubsection{Transformation Ordering}
\label{sec:graphmend-ordering}

\textsc{GraphMend} applies \textsc{[Trap]} first, followed by
\textsc{[Where]}, then \textsc{[Defer]}. Trap
lowering runs first because a branch-local \pyinline{raise} otherwise
prevents control-flow conversion. Once lowered, its tensor assertion
can be conditioned on the enclosing branch predicate. Control-flow
transformation runs next because it may restructure branches, merge
calls, and introduce new return paths. Side effects are deferred last,
after the final control-flow and return structure is known.


\subsection{Semantics Preservation}
\label{sec:graphmend-semantics}
\textsc{GraphMend} preserves the observable behavior of the model:
tensor outputs, visible state, console and log output in
content and order, and raised exceptions with their type and
message. A transformation applies only when every condition in
Table~\ref{tab:graphmend-rules} is established over UniiR; otherwise
the region is left unchanged, so a break that cannot be shown safe
to fix stays in the program.

When its legality conditions hold, each rule does not change the
observable behavior of the model.
\textsc{[Where]} adds only work that raises no exception and writes
nothing visible on the untaken path, and selects the value of the
taken branch. Hoisting can diverge from the original execution in
limited ways, and the legality conditions confine these divergences
to state the model cannot observe. Hoisting a memoized initialization, such as the cached tensor
\pyinline{self.long_inv_freq} in Figure~\ref{fig:phi-4_break}, may
create that tensor earlier than the original program would. This is
safe because the rule accepts such writes only when the written
state stays within the module, the computation is deterministic,
and it cannot raise. The predicate also still selects the same
tensor, so the model output does not change. For inference, the
unselected \pyinline{torch.where} input may also hold NaN or Inf
values, but these never reach the returned tensor. Any branch that
could raise, change observable state, or leak an invalid value on
the untaken path fails the legality check and is not transformed.
\textsc{[Trap]} preserves the guard outcome, since
$\mathit{tensorize}(C)$ has the truth value of \pyinline{C} and the
handler restores the original exception.
\textsc{[Defer]} records argument values at the call site, only on
paths that execute, and flushes them in FIFO order on both normal
and exceptional exits. An exception ends the current FX graph, so
calls recorded before it are replayed at that graph exit before the
exception is re-raised. Calls not yet reached are never recorded. The rules compose in their fixed order, each rechecking
its conditions on the rewritten code.

\begin{table*}[t]
\centering
\caption{
Benchmark suite with graph break counts, causes, fix rates, and cold-start
and steady-state forward pass speedup achieved by fixing graph breaks using
\textsc{GraphMend} over the PyTorch~2 pipeline on RTX
3090, A40 and H100 GPUs.
}
\label{tab:benchmark-merged}
\vspace{-0.3cm}
\fontsize{9.7pt}{10.4pt}\selectfont
\setlength{\tabcolsep}{2.5pt}
\renewcommand{\arraystretch}{1.0}

\begin{tabularx}{\textwidth}{
    @{}
    >{\raggedright\arraybackslash}p{0.23\textwidth}
    >{\centering\arraybackslash}p{0.09\textwidth}
    >{\raggedright\arraybackslash}X
    >{\centering\arraybackslash}p{0.055\textwidth}
    *{6}{>{\centering\arraybackslash}p{0.055\textwidth}}
    @{}
}
\toprule
& & & &
\multicolumn{3}{c}{\textbf{Cold Start}} &
\multicolumn{3}{c}{\textbf{Steady State}} \\
\cmidrule(lr){5-7}
\cmidrule(lr){8-10}

\textbf{Model} &
\textbf{Breaks} &
\textbf{Break Reasons} &
\textbf{Fixed}(\textbf{\%}) &
\textbf{3090} &
\textbf{A40} &
\textbf{H100} &
\textbf{3090} &
\textbf{A40} &
\textbf{H100} \\
\midrule

\multicolumn{10}{@{}l}{
    \textit{Top 30 by downloads (3 of 30 exhibit graph breaks)}
} \\
\midrule

clap-htsat-fused~\cite{clap-model}
& 4
& DS (4)
& 0
& \multicolumn{6}{c@{}}{N/A} \\

chronos-bolt-small~\cite{ansari2024chronos}
& 6
& LC (6)
& 100
& 4.64$\times$
& 4.68$\times$
& 5.12$\times$
& 1.09$\times$
& 1.10$\times$
& 1.15$\times$ \\

whisper-large-v3~\cite{radford2022whisper}
& 3
& LC (3)
& 100
& 3.06$\times$
& 3.34$\times$
& 6.23$\times$
& 1.08$\times$
& 1.07$\times$
& 1.18$\times$ \\

\midrule
\multicolumn{10}{@{}l}{
    \textit{Top 100 trending models (16 of 100 exhibit graph breaks)}
} \\
\midrule

t5-small~\cite{2020t5}
& 3
& LC (3)
& 100
& 3.49$\times$
& 3.40$\times$
& 6.23$\times$
& 1.08$\times$
& 1.06$\times$
& 1.10$\times$ \\

bart-large-cnn~\cite{bart-base}
& 7
& DC (4) + LC (3)
& 100
& 21.07$\times$
& 20.22$\times$
& 24.1$\times$
& 1.13$\times$
& 1.11$\times$
& 1.17$\times$ \\

t5-base~\cite{2020t5}
& 3
& LC (3)
& 100
& 2.27$\times$
& 3.35$\times$
& 3.2$\times$
& 1.09$\times$
& 1.06$\times$
& 1.13$\times$ \\

whisper-small~\cite{radford2022whisper}
& 3
& LC (3)
& 100
& 4.34$\times$
& 5.18$\times$
& 3.42$\times$
& 1.08$\times$
& 1.07$\times$
& 1.09$\times$ \\

inclusively-reformulation-it5~\cite{inc-t5}
& 3
& LC (3)
& 100
& 3.02$\times$
& 2.80$\times$
& 3.98$\times$
& 1.07$\times$
& 1.05$\times$
& 1.07$\times$ \\

whisper-base~\cite{radford2022whisper}
& 3
& LC (3)
& 100
& 2.49$\times$
& 2.61$\times$
& 5.08$\times$
& 1.09$\times$
& 1.10$\times$
& 1.08$\times$ \\

stella-en-400M-v5~\cite{zhang2025jasperstelladistillationsota}
& 4
& DO + DS
& 0
& \multicolumn{6}{c@{}}{N/A} \\

bart-base~\cite{bart-base}
& 7
& DC (4) + LC (3)
& 100
& 11.87$\times$
& 11.39$\times$
& 11.5$\times$
& 1.15$\times$
& 1.14$\times$
& 1.19$\times$ \\

MoLFormer-XL-both-10pct~\cite{MOLformer}
& 5
& VG (5)
& 100
& 7.71$\times$
& 7.35$\times$
& 6.81$\times$
& 1.13$\times$
& 1.11$\times$
& 1.13$\times$ \\

rebel-large~\cite{huguet-cabot-navigli-2021-rebel-relation}
& 7
& DC (4) + LC (3)
& 100
& 19.86$\times$
& 21.90$\times$
& 23.20$\times$
& 1.12$\times$
& 1.14$\times$
& 1.14$\times$ \\

layoutlmv3-base~\cite{huang2022layoutlmv3}
& 2
& LC (2)
& 100
& 23.24$\times$
& 22.59$\times$
& 25.18$\times$
& 1.06$\times$
& 1.07$\times$
& 1.05$\times$ \\

t5-3b~\cite{2020t5}
& 3
& LC (3)
& 100
& 3.01$\times$
& 2.88$\times$
& 2.59$\times$
& 1.08$\times$
& 1.09$\times$
& 1.12$\times$ \\

grounding-dino-tiny~\cite{liu2023grounding}
& 17
& DS (3) + DO (3) + DC (11)
& 58
& 5.20$\times$
& 5.19$\times$
& 5.50$\times$
& 1.08$\times$
& 1.06$\times$
& 1.07$\times$ \\

opus-mt-fr-en~\cite{tiedemann2020opus}
& 6
& DC (3) + LC (3)
& 100
& 13.16$\times$
& 12.80$\times$
& 14.80$\times$
& 1.10$\times$
& 1.12$\times$
& 1.12$\times$ \\

Florence-2-large~\cite{xiao2023florence}
& 7
& DC (7)
& 100
& 1.95$\times$
& 1.55$\times$
& 4.39$\times$
& 1.19$\times$
& 1.21$\times$
& 1.23$\times$ \\

grounding-dino-base~\cite{liu2023grounding}
& 17
& DS (3) + DO (3) + DC (11)
& 58
& 5.17$\times$
& 5.36$\times$
& 5.60$\times$
& 1.06$\times$
& 1.05$\times$
& 1.07$\times$ \\

\midrule
\multicolumn{10}{@{}l}{
    \textit{65 randomly sampled models (8 of 65 exhibit graph breaks)}
} \\
\midrule

Phi-4-mini-instruct~\cite{microsoft_phi4_mini_instruct}
& 5
& DC (5)
& 100
& 3.60$\times$
& 3.90$\times$
& 4.33$\times$
& 1.13$\times$
& 1.14$\times$
& 1.13$\times$ \\

Qwen-Audio-Chat~\cite{Qwen-Audio}
& 2
& DC (2)
& 100
& 2.76$\times$
& 3.27$\times$
& 2.60$\times$
& 1.15$\times$
& 1.16$\times$
& 1.18$\times$ \\

biogpt~\cite{biogpt}
& 2
& DC (2)
& 100
& 2.63$\times$
& 2.86$\times$
& 3.58$\times$
& 1.11$\times$
& 1.14$\times$
& 1.16$\times$ \\

blenderbot-400M-distill~\cite{facebook_blenderbot_400m_distill}
& 3
& LC (3)
& 100
& 3.21$\times$
& 3.85$\times$
& 4.17$\times$
& 1.12$\times$
& 1.09$\times$
& 1.10$\times$ \\

flan-t5-large~\cite{flant5}
& 3
& LC (3)
& 100
& 3.27$\times$
& 3.27$\times$
& 3.51$\times$
& 1.06$\times$
& 1.09$\times$
& 1.12$\times$ \\

tiny-random-PegasusForCausalLM%
~\cite{hf_tiny_random_pegasus_for_causallm}
& 2
& LC (2)
& 100
& 2.72$\times$
& 4.09$\times$
& 5.29$\times$
& 1.34$\times$
& 1.37$\times$
& 1.39$\times$ \\

longformer-base-4096~\cite{Beltagy2020Longformer}
& 5
& LC (3) + TI (2)
& 40
& 2.53$\times$
& 1.95$\times$
& 1.80$\times$
& 1.05$\times$
& 1.07$\times$
& 1.06$\times$ \\

moe-minicpm-x4-base~\cite{babybirdprd_moe_minicpm_x4_base}
& 15
& DS (15)
& 0
& \multicolumn{6}{c@{}}{N/A} \\

\bottomrule
\end{tabularx}


\begin{minipage}{\textwidth}
\fontsize{8.5pt}{9.5pt}\selectfont
\textbf{Break Reason Abbreviations:}
DC\,=\,Data-dependent control flow;\quad
LC\,=\,Logger calls;\quad
DS\,=\,Dynamic shape operator;\quad
DO\,=\,Data-dependent operator;\quad
VG\,=\,\texttt{torch.equal} validation guard;\quad
TI\,=\,\texttt{tensor.item()} call.
\end{minipage}
\vspace{-0.5cm}
\end{table*}


\section{Evaluation}
\label{sec:eval}

We evaluate \textsc{GraphMend} on real-world models that exhibit graph breaks. Our goals are to quantify (i) how often \textsc{GraphMend} removes fixable breaks, (ii) the end-to-end improvement after those fixes, and (iii) what determines the magnitude of improvement.
\vspace{-0.15cm}

\paragraph{\textbf{Experimental Setup}}
We implement GraphMend using the Jaseci framework~\cite{JAC_CAL,jac_OSP}. We compare the original, unmodified model under standard PyTorch 2 with TorchInductor~\cite{pyotrch2} as the backend against the same model with GraphMend enabled. Experiments use NVIDIA RTX 3090, A40, and H100 GPUs under identical configurations. We use PyTorch 2.12, Transformers 4.52.4, CUDA 12.6 and Triton 3.7 versions for all experiments. We capture each fused region into a CUDA Graph~\cite{gray2019cudagraphs}. Each CUDA Graph is replayed with a single launch to isolate GPU execution time when profiling graph breaks. Batch sizes target about 70\% of GPU memory per model; original and fixed models use identical batch sizes and inputs. Text models generate 100 output tokens with greedy decoding.

\vspace{-0.15cm}
\paragraph{\textbf{Benchmark Suite}}
To construct our benchmark suite, we select 195 Hugging Face~\cite{huggingface2025} models covering the top 100 trending models, the top 30 by downloads, and 65 randomly sampled models spanning diverse architectures and task categories. Out of the 195 models, 27 (13.8\%) exhibit graph breaks. We retain all 27 models for evaluation. Table~\ref{tab:benchmark-merged} summarizes the resulting suite, including the number and causes of graph breaks per model. A similar methodology was used in prior work~\cite{pyotrch2}.
\vspace{-0.2cm}

\paragraph{\textbf{Profiling Methodology}}
We profile forward pass execution using PyTorch
Profiler~\cite{pytorch_profiler}. For each model, we evaluate both the original and graph-break-fixed versions over 10 iterations: 1 cold-start iteration followed by 9 warm iterations. We report the forward pass execution latency of the first iteration as the cold-start latency and the mean across the nine subsequent iterations as the steady-state latency. The profiler records CPU, CUDA, and operator-level events, which we analyze using Chrome's trace viewer.
 
\vspace{-0.25cm}

\subsection{\textsc{GraphMend}'s Ability to Fix Graph Breaks}
\label{subsec:fix_breaks}

We evaluate \textsc{GraphMend}'s effectiveness in fixing graph breaks across the full benchmark suite of 27 models.
As shown in Table~\ref{tab:benchmark-merged}, \textsc{GraphMend} fully eliminates all graph breaks in 21 models and partially fixes 3 others. Only 3 models remain entirely unfixed. In total, \textsc{GraphMend} removes 107 out of 147 graph breaks across the suite (73\%).

Graph breaks due to data-dependent control flow (\pyinline{if}/\pyinline{else} on tensor values), logger/print side effects, and validation guards are all resolved by the three transformations. The remaining unfixed breaks are caused by \pyinline{tensor.item()} calls and dynamic shape operators, which cannot be resolved through code transformations because they require runtime values that have no tensor-level equivalent.


\vspace{-0.2cm}

\paragraph{Output correctness validation.}
We compare the original and \textsc{GraphMend} transformed versions of all 24 fully or partially fixable models using identical inputs. All FP32 outputs are bit-identical and greedy-decoded token sequences are identical for every generative model. These results provide empirical evidence that \textsc{GraphMend} preserves numerical outputs on the evaluated workloads.

For the performance analysis that follows, we report results on the 24 models where graph breaks were fully or partially eliminated.

\subsection{Cold Start Speedup}
\label{subsec:cold_start}

We measure the cold-start speedup as the ratio of original to fixed forward-pass latency during the first forward pass after compilation, when CUDA graphs must be recorded and cached.
 
Table~\ref{tab:benchmark-merged} shows results on RTX~3090, A40 and H100 GPUs across all 24 models. We observe speedups from 2$\times$ to 26$\times$ (5$\times$ on average). Every model shows substantial improvement, regardless of the break type. This pattern is consistent across all three GPUs, confirming that cold-start overhead is a software-level bottleneck in the compilation pipeline, independent of GPU hardware.

At cold start, each break triggers Dynamo tracing and backend compilation of the next FX subgraph inside the measured forward pass, whereas a break-free forward pass completes both tracing and compilation before execution begins. This is why models with more graph breaks tend to achieve higher cold-start speedups; a detailed trace-level analysis appears in~\S~\ref{subsec:case study}.

The bart-large-cnn model illustrates the magnitude of this
overhead. Its 7 graph breaks split the forward pass into 8 FX
subgraphs, each compiled when execution first reaches it. Fixing
them reduces the first forward-pass latency from 12.18~s to
580~ms, a 21$\times$ speedup. In the original trace, backend
compilation across the subgraphs accounts for 80\% of this
latency, Dynamo graph capture and device synchronization for
another 17\%, and GPU execution for the remainder.

\begin{figure*}[t]
    \centering
    \begin{subfigure}[t]{0.70\textwidth}
        \centering
        \includegraphics[width=\linewidth]{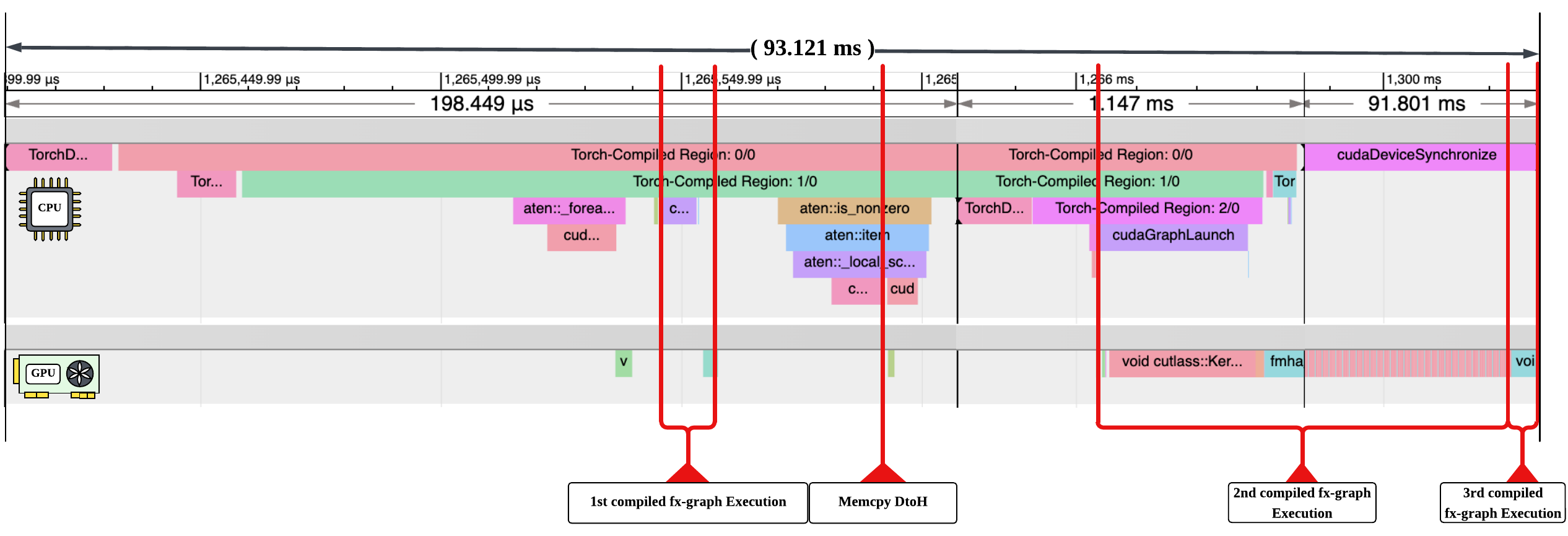}
        \caption{}
        \label{fig:original_breakdown}
    \end{subfigure}
    \vspace{-0.05cm}
    \begin{subfigure}[t]{0.70\textwidth}
        \centering
        \includegraphics[width=\linewidth]{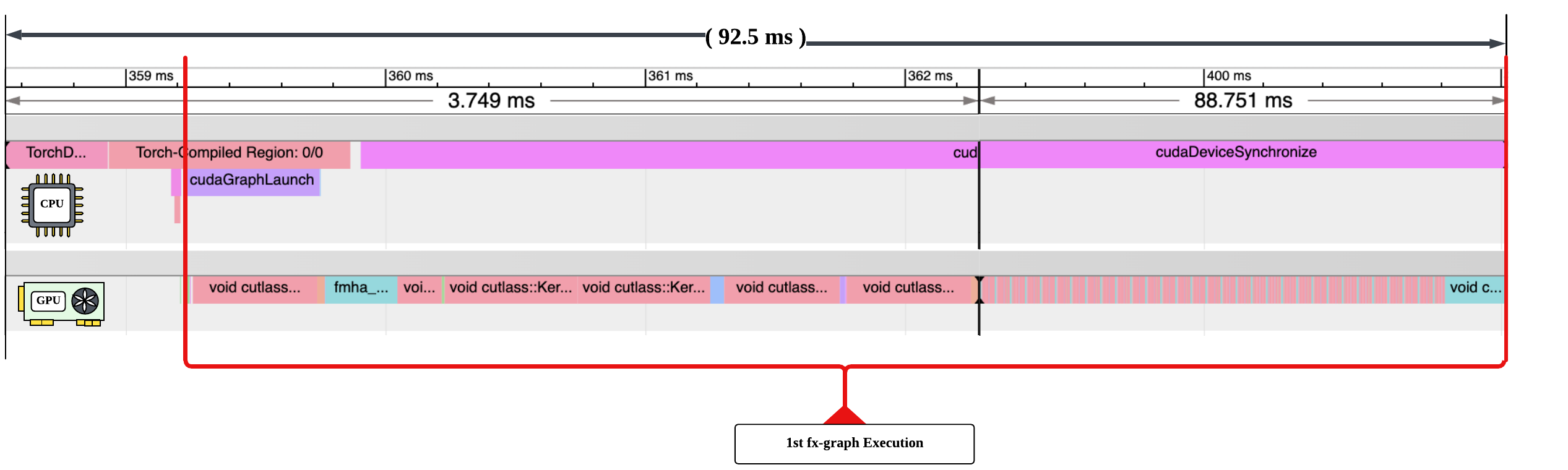}
        \caption{}
        \label{fig:fixed_breakdown}
    \end{subfigure}
    \vspace{-0.2cm}
    \caption{Profiler traces for one warm-run iteration of Qwen-Audio-Chat on an A40 GPU. (a)~The \emph{original} model is split into three separate CUDA graphs due to two graph breaks. (b)~The \emph{fixed} model executes as a single continuous CUDA graph with no GPU idle gaps.}
    \label{fig:traces}
    \vspace{-0.3cm}
\end{figure*}

\begin{figure}[t]
    \centering
    \begin{subfigure}[t]{0.23\textwidth}
        \centering
        \includegraphics[width=\linewidth]{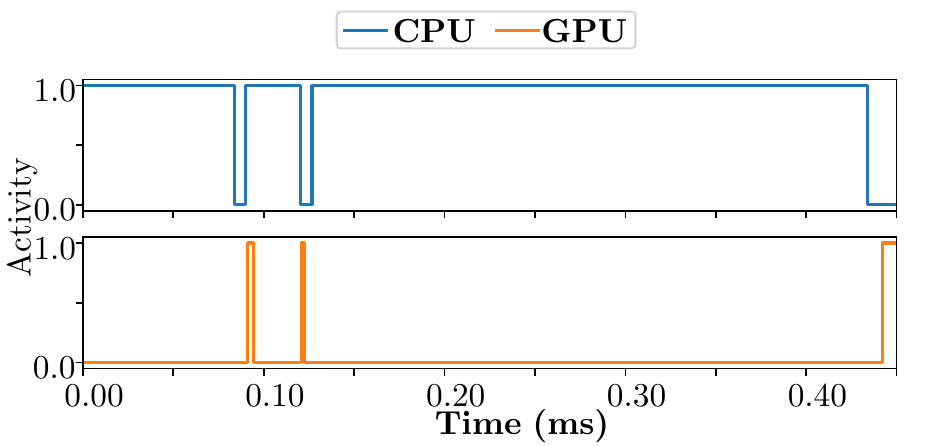}
        \caption{Steady State -- Original Model}
        \label{fig:original_activity}
    \end{subfigure}
    \begin{subfigure}[t]{0.23\textwidth}
        \centering
        \includegraphics[width=\linewidth]{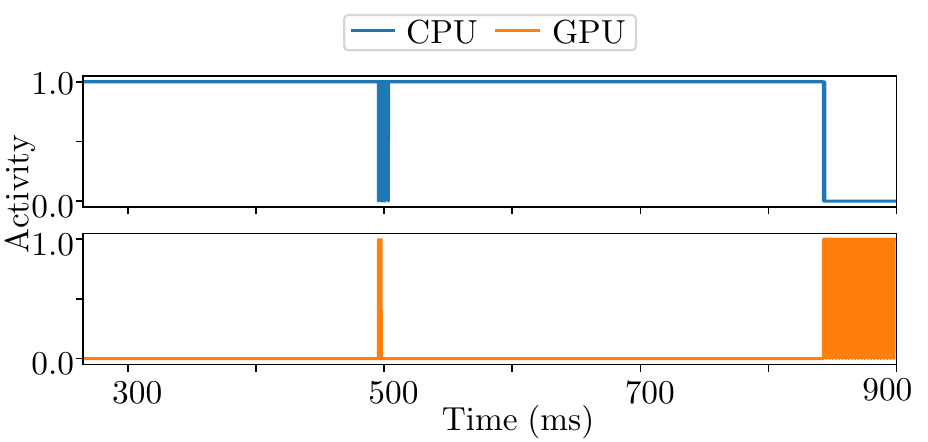}
        \caption{Cold Start -- Original Model}
        \label{fig:original_cold_activity}
    \end{subfigure}
    \hfill
    \begin{subfigure}[t]{0.23\textwidth}
        \centering
        \includegraphics[width=\linewidth]{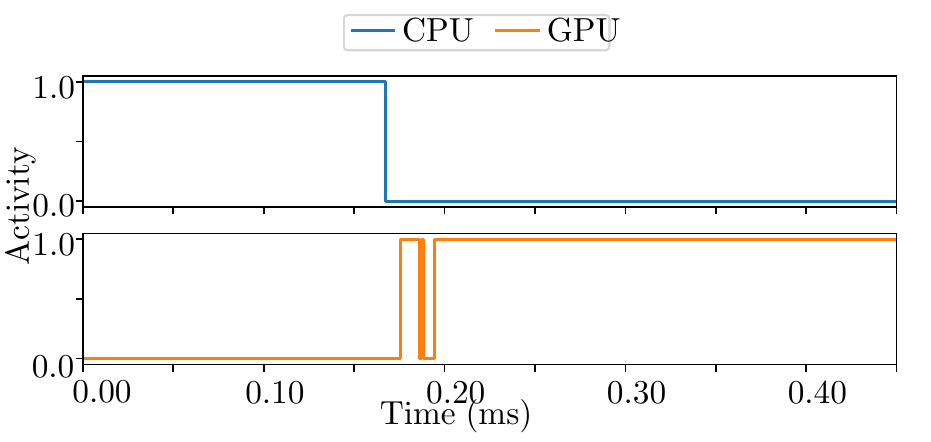}
        \caption{Steady State -- Fixed Model}
        \label{fig:fixed_activity}
    \end{subfigure}
    \begin{subfigure}[t]{0.23\textwidth}
        \centering
        \includegraphics[width=\linewidth]{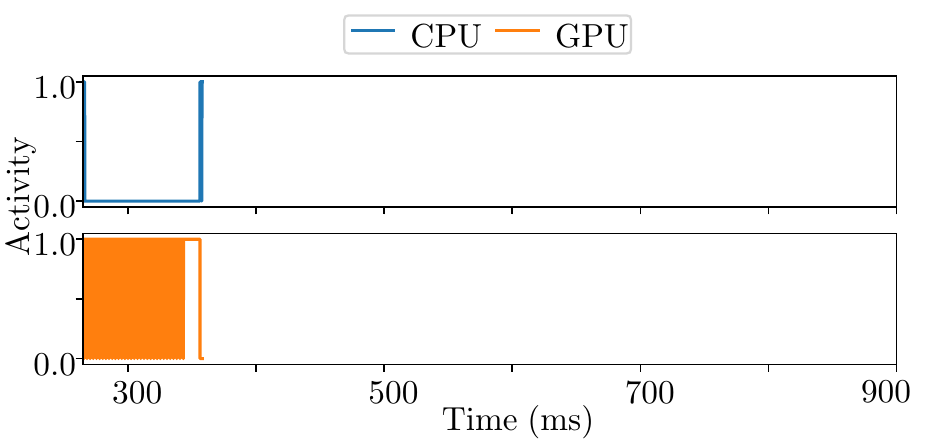}
        \caption{Cold Start -- Fixed Model}
        \label{fig:fixed_cold_activity}
    \end{subfigure}
    \caption{CPU/GPU activity time traces for Qwen-Audio-Chat model run on A40 GPU. The original model shows GPU idle periods between graph executions due to D2H memcpy and CPU fallback vs. continuous GPU activity of fixed model.}
    \label{fig:activity}
    \vspace{-0.3cm}
\end{figure}

\subsection{Steady-State Forward Pass Speedup}
\label{subsec:warm}

We measure the steady-state speedup as the ratio of original to fixed forward-pass latency during runs after the first iteration, when CUDA graphs have already been compiled and cached for replay.
 
Table~\ref{tab:benchmark-merged} shows steady-state speedup on RTX~3090, A40 and H100 GPUs. All models show steady-state speedups, ranging from 1.05$\times$ to 1.39$\times$. Unlike the cold-start results, the magnitude of steady-state improvement varies across models depending on the type of graph break.
 

Models with data-dependent control flow breaks show the largest gains. This is because each such break forces a device synchronization and a D2H memcpy so that Python can evaluate the tensor-dependent condition, leaving the GPU idle during this sequence (\S~\ref{subsec:case study}). Fixing these breaks allows the forward pass to execute as a single continuous CUDA graph without interruption. In addition, eliminating graph breaks allows TorchInductor to fuse kernels across former break boundaries, further reducing the number of kernel launches and improving steady-state performance.

Since \textsc{[Where]} executes both hoisted branches, we quantify the cost of predication by comparing three versions of Phi-4-mini-instruct: the original graph-breaking code, a manually fixed single-branch version, and \textsc{GraphMend}'s
\pyinline{torch.where} rewrite. Their steady-state GPU busy times are 100.407~ms, 99.604~ms, and 99.643~ms, respectively. The nearly identical single-branch and \textsc{GraphMend} times show no meaningful extra GPU work from executing both branches. Across the full suite, no rewrite caused a slowdown: every steady-state result in Table~\ref{tab:benchmark-merged} is at least 1.05$\times$ faster. 
 
Models with logger-only breaks show smaller but consistent gains. In the models we observe, the logger calls involve only pure Python strings (e.g., \pyinline{logger.warning\_once("deprecated...")}), so no tensor values need to be transferred to the CPU. However, if a logger call were to involve a tensor value (e.g., logging \pyinline{x.max()}), it would force a D2H memcpy similar to data-dependent breaks. Even without D2H synchronization, eliminating logger breaks still reduces the number of separate CUDA graph launches and enables kernel fusion across the former break boundaries.
 
Notably, tiny-random-PegasusForCausalLM shows 1.39$\times$ speedup. Since this model is very small (64~operators), the fixed costs from graph breaks take up a large portion of the runtime, so removing them gives a large speedup per forward pass. This result is consistent across all GPUs.

\subsection{Throughput Improvement}
\label{subsec:throughput}
\begin{figure}[h]
    \centering
    \includegraphics[width=0.96\linewidth]{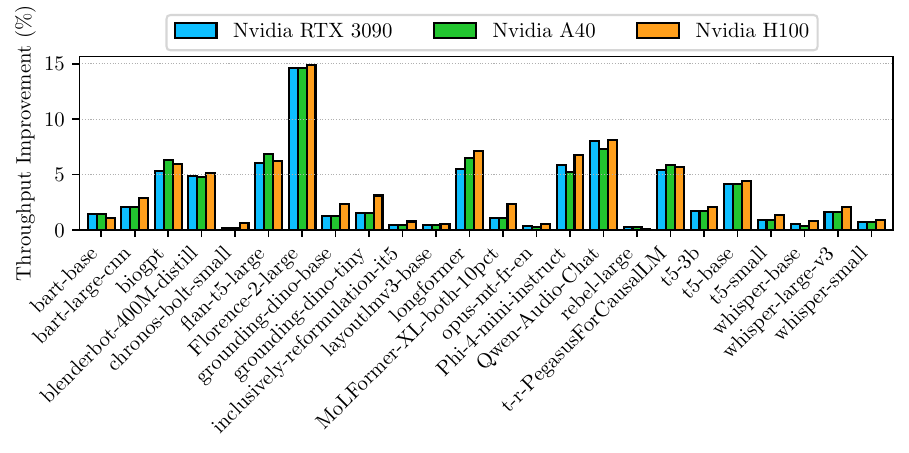}
    \caption{Throughput gain across the benchmark suite.}
    \label{fig:throughput}
\end{figure}
We measure the throughput improvement as the percentage increase in inference throughput (e.g., tokens per second for generative models, samples per second for encoders) after fixing graph breaks. As shown in Figure~\ref{fig:throughput}, we observe up to 15\% throughput improvement across models. Results are consistent across all RTX~3090, A40 and H100 GPUs.

Throughput gains are smaller than steady-state forward pass gains because end-to-end inference involves more than just the forward pass. For autoregressive models, each output token requires a separate decode step that is bounded by memory bandwidth rather than compute. On top of that, tokenization, sampling, and CPU-side scheduling all add time that \textsc{GraphMend} does not affect. By Amdahl's law, the end-to-end throughput improvement is limited by the fraction of time spent in the forward pass.

This is why Florence-2-large shows the highest throughput gain (15\%): as a vision-language model, its forward pass is a single non-autoregressive pass through the vision encoder, making it a larger fraction of total inference time. In contrast, autoregressive models like Qwen-Audio-Chat (8\%) and Phi-4-mini-instruct (6\%) spend most of their time in the decode loop, so even with a faster forward pass, throughput improves less. The variation between these two models reflects graph break location: the break in Qwen-Audio-Chat occurs mid-function where its impact is greater, while the break in Phi-4-mini-instruct occurs early.



\vspace{-0.1cm}
\subsection{Overhead Analysis of Graph Breaks}
\label{subsec:case study}

Sections~\ref{subsec:cold_start} and~\ref{subsec:warm} attributed the speedups to CUDA graph recording overhead and CPU--GPU synchronization. In this section, we verify this attribution at the trace level by analyzing profiler traces of Qwen Audio Chat on an A40 GPU, which contains two graph breaks caused by data-dependent control flow. For a consistent comparison, all activity plots (Figures~\ref{fig:original_activity}--\ref{fig:fixed_cold_activity}) show the time interval between the first and second CUDA graph executions in the original model; the same interval is used for the fixed model.

\vspace{-0.13cm}

\paragraph{\textbf{Steady-state run overhead analysis}}
Figure~\ref{fig:original_breakdown} shows the profiler trace for the original model. The forward pass is split into 3 CUDA graphs due to the 2 graph breaks. Between the first and second FX graph executions, we observe: (1)~a \texttt{cudaStreamSynchronize} that blocks the CPU until the first graph finishes, (2)~a device-to-host memcpy that transfers the tensor value needed for the Python conditional, and (3)~a CPU-side evaluation of the \pyinline{if} condition in eager mode. During this entire sequence, the GPU is idle, as Figure~\ref{fig:original_activity} confirms. After applying \textsc{GraphMend}, the trace (Figure~\ref{fig:fixed_breakdown}) shows one continuous CUDA graph with no gaps. The synchronization, memcpy, and eager-mode evaluation are all gone. Figure~\ref{fig:fixed_activity} confirms that GPU activity remains uninterrupted throughout the forward pass.

\paragraph{\textbf{Cold run overhead analysis}} 
The cold-run traces reveal additional overhead not present in steady state. In the original model (Figure~\ref{fig:original_cold_activity}), Dynamo traces and the backend compiles only the first subgraph before forward pass execution begins. When forward pass execution reaches a break, it falls back to eager mode, and Dynamo tracing and backend compilation of the next subgraph run in the middle of the forward pass while the GPU sits idle. This repeats at every break, so GPU idle time far exceeds GPU active time during the cold run. In the fixed model (Figure~\ref{fig:fixed_cold_activity}), Dynamo traces and compiles the entire forward pass as one graph before forward pass execution begins, so the execution itself contains no compilation work. The total GPU idle time in the fixed cold run is shorter than the original model run.



 

\subsection{Full-Graph Capture for Serving Frameworks}
\label{subsec:fullgraph}
Serving frameworks place strict requirements on model code: vLLM requires that model code be capturable into a full graph via Dynamo (\pyinline{torch.compile(fullgraph=True)}) to use its compilation and CUDA Graph replay path~\cite{vllm-debug-compile}, and SGLang's piecewise CUDA graph similarly relies on \pyinline{torch.compile} tracing of the model's forward pass~\cite{sglang-pcg}. A model with even one graph break fails this requirement~\cite{vllm-debug-compile, pytorch-fullgraph}. For each of the 21 fully fixed models in Table~\ref{tab:benchmark-merged}, we attempt \pyinline{torch.compile(fullgraph=True)} on the original and the \textsc{GraphMend}-fixed model, using \pyinline{backend="eager"} to isolate Dynamo's graph capture from backend compilation. All original models fail, whereas all fixed models succeed, confirming the elimination of graph breaks.

We ran the fixed Phi-4-mini model in vLLM through its Transformers backend~\cite{vllm-transformers-backend}. The fix clears the graph break, but vLLM then rejects the model because the Hugging Face rotary embedding mutates the \pyinline{inv_freq} buffer inside \pyinline{forward}, which vLLM forbids under CUDA Graph replay. This is a runtime constraint unrelated to graph breaks; vLLM's own hand-written models avoid it by precomputing rotary caches at initialization. After manually applying this precomputation, the fixed model runs end-to-end in vLLM with coherent output. \textsc{GraphMend} thus removes the graph-capture barrier to serving these models directly, leaving only framework-specific runtime constraints.



\subsection{\textsc{GraphMend} Compilation Overhead}

\begin{figure}[t]
\centering
\includegraphics[width=0.8\linewidth]{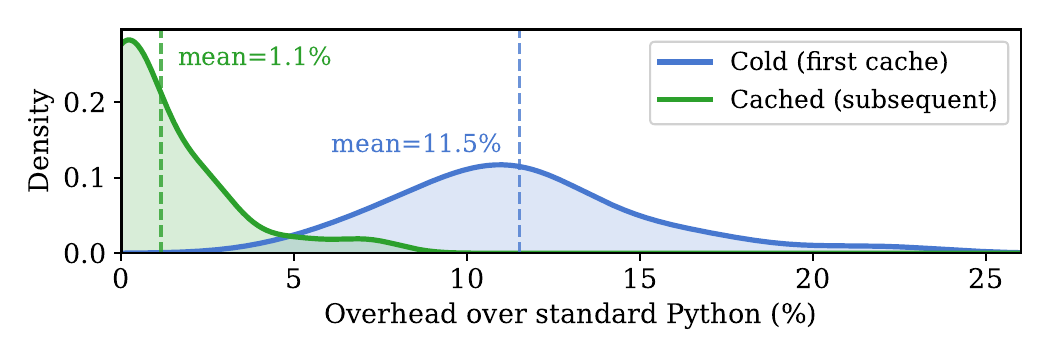}
\vspace{-0.4cm}
\caption{Distribution of \textsc{GraphMend}'s compilation overhead relative to standard Python across 24 models.}
\label{fig:overhead}
\vspace{-0.5cm}
\end{figure}

To measure the overhead of \textsc{GraphMend}'s compiler analysis and
transformations, we compare the end-to-end wall-clock time of each model
executed through \textsc{GraphMend} against the same model script running
under standard Python. Both configurations execute the same script with
identical inputs, differing only in whether \textsc{GraphMend}'s
compilation pipeline is applied before execution. We evaluate a cold run,
in which the compiler cache is empty and all passes execute, and a cached
run, in which the generated artifacts are reused. As shown in
Figure~\ref{fig:overhead}, the cold run incurs a mean overhead of 11.5\%,
while the cached run incurs only 1.1\%. The cold overhead corresponds to a
fixed, one-time compilation cost of roughly 0.5\,s per model and therefore
diminishes for longer-running workloads. The distribution also becomes
tightly concentrated near zero after caching, showing that most of
\textsc{GraphMend}'s compilation cost is incurred only during the first
execution and is largely amortized in subsequent runs.





\definecolor{lightgreen}{RGB}{204,255,204}
\definecolor{lightred}{RGB}{255,204,204}
\definecolor{lightyellow}{RGB}{255,255,204}
\definecolor{orange}{RGB}{255,229,180}

\section{Related Work}

\begin{table}[h]
\centering
\vspace{-0.2cm}
\caption{Comparison of \textsc{GraphMend} with related transformation frameworks. $\checkmark$ denotes full support and $\sim$ denotes partial support.}
\vspace{-0.3cm}
\label{tab:graphmend-comparison}
\resizebox{\columnwidth}{!}{
\begin{tabular}{|p{3.9cm}|c|c|c|c|}
\hline
\textbf{Feature} & \textbf{AutoGraph} & \textbf{MagPy} & \textbf{TensorSSA} & \textbf{GraphMend} \\
& \cite{autograph} & \cite{MagPy} & \cite{TensoSSA} & \\
\hline
Supports standard PyTorch~2 pipeline & \cellcolor{lightred} & \cellcolor{lightred} & \cellcolor{lightred} & \cellcolor{lightgreen}{$\checkmark$} \\
\hline
Leverage source-level semantics for analysis & \cellcolor{lightgreen}{$\checkmark$} & \cellcolor{lightgreen}{$\checkmark$} & \cellcolor{lightred} & \cellcolor{lightgreen}{$\checkmark$} \\
\hline
Performs code transformations before runtime & \cellcolor{lightgreen}{$\checkmark$} & \cellcolor{lightred} & \cellcolor{lightgreen}{$\checkmark$} & \cellcolor{lightgreen}{$\checkmark$} \\
\hline
Resolves data-dependent control flow statically, without runtime dispatch & \cellcolor{lightyellow}{$\sim$} & \cellcolor{lightyellow}{$\sim$} & \cellcolor{lightred} & \cellcolor{lightgreen}{$\checkmark$} \\
\hline
Defers side effects (\texttt{print}/logger) past the compiled region & \cellcolor{lightyellow}{$\sim$} & \cellcolor{lightred} & \cellcolor{lightred} & \cellcolor{lightgreen}{$\checkmark$} \\
\hline
Handles \texttt{raise}-based validation guards & \cellcolor{lightred} & \cellcolor{lightred} & \cellcolor{lightred} & \cellcolor{lightgreen}{$\checkmark$} \\
\hline
Preserves exact Python execution semantics & \cellcolor{lightred} & \cellcolor{lightgreen}{$\checkmark$} & \cellcolor{lightgreen}{$\checkmark$} & \cellcolor{lightgreen}{$\checkmark$} \\
\hline
\end{tabular}
}
\end{table}

PyTorch has long sought to balance performance with Python's flexibility. TorchScript enabled ahead-of-time graph capture but struggled with data-dependent control flow~\cite{pytorch_jit_script,pytorch_jit_trace}, while PyTorch~2's TorchDynamo~\cite{pyotrch2,pytorch} still introduces graph breaks. \textsc{GraphMend} addresses these breaks by transforming the Python AST before execution while preserving the standard PyTorch~2 compilation pipeline.

AutoGraph~\cite{autograph} also rewrites Python control flow at the source level. However, it targets TensorFlow, where Python conditionals cannot directly execute in a graph, and differs from \textsc{GraphMend} in several ways. First, it addresses only data-dependent control flow, one of \textsc{GraphMend}'s three transformation categories, and converts conditionals broadly into runtime dispatch code, whereas \textsc{GraphMend} uses static use-def analysis to rewrite only the predicates that depend on tensor data. Second, AutoGraph converts \texttt{print} to \texttt{tf.print} but leaves it at its graph position, where printing GPU tensors can synchronize; \textsc{GraphMend} defers \texttt{print} and logger effects past the compiled region and additionally handles \texttt{raise}-based validation guards. Third, AutoGraph translates programs into TensorFlow graph semantics, whereas \textsc{GraphMend} must eliminate Dynamo's eager fallback while preserving full Python behavior, including outputs, side effects, state updates, and exceptions.

MagPy~\cite{MagPy} replaces TorchDynamo with a runtime capture mechanism that records one execution, guards its assumptions, and recompiles when they fail. It does not remove data-dependent control flow; it moves the break into runtime as an \texttt{If} node, which requires backend support for graph-level control flow and otherwise falls back to eager execution. It also leaves \texttt{print} and logger calls and \texttt{raise}-based validation guards unaddressed. \textsc{GraphMend} instead rewrites break-causing constructs at the source, so the unmodified TorchDynamo and backend stack capture and optimize a complete graph.

TensorSSA~\cite{TensoSSA} removes tensor mutation through a custom SSA-based IR on the TorchScript path, operating only after graph capture; it is therefore complementary, since \textsc{GraphMend} enlarges the graphs available for post-capture optimization while applying SSA-inspired rewrites directly on the Python AST. More broadly, graph optimizers~\cite{openxla2023,chen2018tvm,lattner2020mlir,rotem2018glow,jia2019taso,onnx2018} assume a complete graph already exists; \textsc{GraphMend} restructures the source to provide one. Since these systems target different compiler stacks, a direct empirical comparison would conflate framework differences with the effect of graph-break repair; our evaluation therefore isolates that effect against the standard PyTorch~2 pipeline.
\section{Conclusion}



We present \textsc{GraphMend}, a compiler technique for PyTorch that systematically eliminates common FX graph breaks. By restructuring code constructs such as data-dependent control flow, Python side effects, and validation guards, it reduces fragmentation in the compilation pipeline and enables TorchDynamo to capture larger, unified graphs. Our evaluation across 27 Hugging Face models shows that we remove 73\% of all graph breaks, achieving up to 26$\times$ cold-start speedup on NVIDIA GPUs.
\appendix
\section{Artifact Appendix}

\subsection{Abstract}

The GraphMend artifact contains the compiler implementation and evaluation
scripts for the 27 models in Table~\ref{tab:benchmark-merged}. It supports three claims from \S~\ref{sec:eval}:

\begin{itemize}
    \item \textbf{C1:} GraphMend fixes graph breaks while preserving the outputs of the original model.
    \item \textbf{C2:} Models for which GraphMend eliminates all graph breaks can
be captured using \texttt{torch.compile} with \texttt{fullgraph=True},
satisfying a key prerequisite for modern serving frameworks that rely on
full-graph capture.
    \item \textbf{C3:} Fixing graph breaks improves PyTorch 2 execution
performance by reducing cold-start and steady-state forward-pass latency.
\end{itemize}

C1 and C2 can run on CPU or GPU, while C3 requires an NVIDIA GPU.
The artifact is available at
\url{https://github.com/savini98/CGO_AE_GraphMend}.

\subsection{Artifact Check-list (Meta-information)}

\begin{itemize}
    \item \textbf{Program:} GraphMend implemented in the Jaseci/Jac compiler.
    \item \textbf{Models:} The 27 Hugging Face models in Table~\ref{tab:benchmark-merged}.
    \item \textbf{Data set:} None. Original and fixed models use identical
    inputs.
    \item \textbf{Software:} Docker, Python 3.13, PyTorch 2.12.1,
    Transformers 4.52.4; C3 additionally uses CUDA 12.6 and Triton 3.7.
    \item \textbf{Hardware:} C1 and C2 require at least 20\,GB RAM and
60\,GB of free disk space; 100\,GB of free disk space is recommended.
C3 requires the same resources in addition to an NVIDIA GPU with CUDA
support. The performance results reported in \S~\ref{sec:eval} were collected on
RTX 3090, A40, and H100 GPUs.
    \item \textbf{Metrics:} graph-break count, output equivalence,
    full-graph capture, cold-start latency, and steady-state latency.
   \item \textbf{Time:} C1 and C2 take approximately 10--12 hours.
C3 takes approximately 5 hours for the 10-model subset and around
12 hours for the complete model set on one GPU.
\item \textbf{Availability:}
\url{https://github.com/savini98/CGO_AE_GraphMend};
archival DOI: \href{https://doi.org/10.5281/zenodo.22242705}
{zenodo.22242705}.
\end{itemize}

\subsection{Description}

\paragraph{Hardware and software.}
Docker is the recommended execution path and pins the software environment.
At least 20\,GB RAM and 60\,GB of free disk space are required, with
100\,GB of free disk space recommended. C1 and C2 can execute on CPU or
GPU. C3 can be executed on any NVIDIA GPU with CUDA support. The performance
results reported in \S~\ref{sec:eval} were collected on RTX 3090, A40, and H100 GPUs.
These specific GPUs are only required to reproduce the corresponding
hardware configurations from the paper; they are not required to run C3.

\paragraph{Models and inputs.}
The artifact evaluates the same 27 models reported in Table~\ref{tab:benchmark-merged}. The original
and GraphMend-fixed versions receive identical inputs, allowing graph-break
counts and model outputs to be compared directly.

\subsection{Installation}

Clone the repository and build the Docker image:

\begin{lstlisting}[style=artifactbash]
git clone --recurse-submodules https://github.com/savini98/CGO_AE_GraphMend
cd CGO_AE_GraphMend
docker build -f artifact/Dockerfile -t graphmend .
\end{lstlisting}

\subsection{Experiment Workflow}

\paragraph{C1: Fixing Graph Breaks.}
Run:
\begin{lstlisting}[style=artifactbash]
docker run --rm graphmend c1
\end{lstlisting}
For each model, C1 reports graph breaks before and after GraphMend and
compares the original and fixed outputs. Correctness is evaluated together with graph-break fixing
because a transformation is considered a valid fix only when
model outputs are preserved.

\paragraph{C2: Full-Graph Capture.}
Run:
\begin{lstlisting}[style=artifactbash]
docker run --rm graphmend c2
\end{lstlisting}
C2 applies \texttt{torch.compile(fullgraph=True)} to the original and fixed
models. As in \S~\ref{subsec:fullgraph}, the eager backend is used to isolate TorchDynamo graph
capture from backend compilation.

\paragraph{C3: GPU Performance.}
Run:
\begin{lstlisting}[style=artifactbash]
docker run --rm --gpus all graphmend c3
\end{lstlisting}
C3 follows the profiling methodology of \S~\ref{sec:eval}. Each model executes 10
forward-pass iterations, with the first iteration reported as cold start and
the following nine used for steady-state latency. Five profiler traces are also generated per model.

The base C3 command evaluates a representative 10-model
subset and takes around 5 hours. This uses the same
profiling procedure as the complete evaluation while reducing
evaluation time. To run the complete model set:

\begin{lstlisting}[style=artifactbash]
docker run --rm --gpus all graphmend c3 --full
\end{lstlisting}

The full C3 workflow takes around 12 hours on one GPU.

\subsection{Evaluation and Expected Results}

\paragraph{C1.}
C1 reports the number of graph breaks before and after GraphMend and compares
the outputs of the original and fixed models. The resulting graph-break
elimination and correctness results can be compared against Table~\ref{tab:benchmark-merged} and
\S~\ref{subsec:fix_breaks}. Known configuration-dependent differences in individual break counts
are documented in \texttt{artifact/README.md}.

\paragraph{C2.}
C2 reports whether the original and fixed models can be captured using
\texttt{torch.compile} with \texttt{fullgraph=True}. The results can be
compared against the full-graph capture results in \S~\ref{subsec:fullgraph}.

\paragraph{C3.}
C3 reports cold-start and steady-state forward-pass latency for the original
and fixed models. The resulting performance trends can be compared against
the results in Table~\ref{tab:benchmark-merged}. Exact latency and speedup values may vary with the GPU
and system configuration.

\subsection{Experiment Customization}

C1 can be run on individual models by appending the model
name to the command. C3 runs the 10-model subset with the base
command; \texttt{--full} runs the complete evaluation set.
Known configuration-dependent differences are documented in
\texttt{artifact/README.md}.



\bibliographystyle{ACM-Reference-Format}
\bibliography{sample-base}

@inproceedings{pyotrch2,
author = {Ansel, Jason and Yang, Edward and He, Horace and Gimelshein, Natalia and Jain, Animesh and Voznesensky, Michael and Bao, Bin and Bell, Peter and Berard, David and Burovski, Evgeni and Chauhan, Geeta and Chourdia, Anjali and Constable, Will and Desmaison, Alban and DeVito, Zachary and Ellison, Elias and Feng, Will and Gong, Jiong and Gschwind, Michael and Hirsh, Brian and Huang, Sherlock and Kalambarkar, Kshiteej and Kirsch, Laurent and Lazos, Michael and Lezcano, Mario and Liang, Yanbo and Liang, Jason and Lu, Yinghai and Luk, C. K. and Maher, Bert and Pan, Yunjie and Puhrsch, Christian and Reso, Matthias and Saroufim, Mark and Siraichi, Marcos Yukio and Suk, Helen and Zhang, Shunting and Suo, Michael and Tillet, Phil and Zhao, Xu and Wang, Eikan and Zhou, Keren and Zou, Richard and Wang, Xiaodong and Mathews, Ajit and Wen, William and Chanan, Gregory and Wu, Peng and Chintala, Soumith},
title = {PyTorch 2: Faster Machine Learning Through Dynamic Python Bytecode Transformation and Graph Compilation},
year = {2024},
isbn = {9798400703850},
publisher = {Association for Computing Machinery},
address = {New York, NY, USA},
url = {https://doi.org/10.1145/3620665.3640366},
doi = {10.1145/3620665.3640366},
booktitle = {Proceedings of the 29th ACM International Conference on Architectural Support for Programming Languages and Operating Systems, Volume 2},
pages = {929–947},
numpages = {19},
location = {La Jolla, CA, USA},
series = {ASPLOS '24}
}

@inbook{pytorch,
author = {Paszke, Adam and Gross, Sam and Massa, Francisco and Lerer, Adam and Bradbury, James and Chanan, Gregory and Killeen, Trevor and Lin, Zeming and Gimelshein, Natalia and Antiga, Luca and Desmaison, Alban and K\"{o}pf, Andreas and Yang, Edward and DeVito, Zach and Raison, Martin and Tejani, Alykhan and Chilamkurthy, Sasank and Steiner, Benoit and Fang, Lu and Bai, Junjie and Chintala, Soumith},
title = {PyTorch: an imperative style, high-performance deep learning library},
year = {2019},
publisher = {Curran Associates Inc.},
address = {Red Hook, NY, USA},
booktitle = {Proceedings of the 33rd International Conference on Neural Information Processing Systems},
articleno = {721},
numpages = {12}
}

@inproceedings{GRAPE,
author = {Zheng, Bojian and Yu, Cody Hao and Wang, Jie and Ding, Yaoyao and Liu, Yizhi and Wang, Yida and Pekhimenko, Gennady},
title = {Grape: Practical and Efficient Graphed Execution for Dynamic Deep Neural Networks on GPUs},
year = {2023},
isbn = {9798400703294},
publisher = {Association for Computing Machinery},
address = {New York, NY, USA},
url = {https://doi.org/10.1145/3613424.3614248},
doi = {10.1145/3613424.3614248},
booktitle = {Proceedings of the 56th Annual IEEE/ACM International Symposium on Microarchitecture},
pages = {1364–1380},
numpages = {17},
location = {Toronto, ON, Canada},
series = {MICRO '23}
}

@ARTICLE{JAC_CAL,
  author={Mars, Jason and Kang, Yiping and Daynauth, Roland and Li, Baichuan and Mahendra, Ashish and Flautner, Krisztian and Tang, Lingjia},
  journal={IEEE Computer Architecture Letters}, 
  title={The Jaseci Programming Paradigm and Runtime Stack: Building Scale-Out Production Applications Easy and Fast}, 
  year={2023},
  volume={22},
  number={2},
  pages={101-104},
  doi={10.1109/LCA.2023.3274038}}

@misc{jac_OSP,
      title={Extending Data Spatial Semantics for Scale Agnostic Programming}, 
      author={Jason Mars},
      year={2025},
      eprint={2504.03109},
      archivePrefix={arXiv},
      primaryClass={cs.PL},
      url={https://arxiv.org/abs/2504.03109}, 
}

@misc{mtp,
      title={MTP: A Meaning-Typed Language Abstraction for AI-Integrated Programming}, 
      author={Jayanaka L. Dantanarayana and Yiping Kang and Kugesan Sivasothynathan and Christopher Clarke and Baichuan Li and Savini Kashmira and Krisztian Flautner and Lingjia Tang and Jason Mars},
      year={2025},
      eprint={2405.08965},
      archivePrefix={arXiv},
      primaryClass={cs.PL},
      url={https://arxiv.org/abs/2405.08965}, 
}

@misc{torch.fx,
      title={Torch.fx: Practical Program Capture and Transformation for Deep Learning in Python}, 
      author={James K. Reed and Zachary DeVito and Horace He and Ansley Ussery and Jason Ansel},
      year={2022},
      eprint={2112.08429},
      archivePrefix={arXiv},
      primaryClass={cs.LG},
      url={https://arxiv.org/abs/2112.08429}, 
}

@misc{pytorch_jit_trace,
  title        = {{torch.jit.trace} — PyTorch Documentation},
  author       = {{PyTorch Team}},
  howpublished = {\url{https://docs.pytorch.org/docs/stable/generated/torch.jit.trace.html}},
  note         = {Accessed: 2025-09-08},
  year         = {2025}
}

@misc{pytorch_jit_script,
  title        = {{torch.jit.script} — PyTorch Documentation},
  author       = {{PyTorch-TorchScript Team}},
  howpublished = {\url{https://docs.pytorch.org/docs/stable/generated/torch.jit.script.html}},
  note         = {Accessed: 2025-09-08},
  year         = {2025}
}

@misc{ghosh2025pygraphrobustcompilersupport,
      title={PyGraph: Robust Compiler Support for CUDA Graphs in PyTorch}, 
      author={Abhishek Ghosh and Ajay Nayak and Ashish Panwar and Arkaprava Basu},
      year={2025},
      eprint={2503.19779},
      archivePrefix={arXiv},
      primaryClass={cs.LG},
      url={https://arxiv.org/abs/2503.19779}, 
}

@misc{pytorch_compiler_troubleshooting,
  title        = {Torch Compile Troubleshooting — PyTorch Documentation},
  author       = {{PyTorch Contributors}},
  howpublished = {\url{https://docs.pytorch.org/docs/stable/torch.compiler_troubleshooting.html}},
  note         = {Accessed: 2025-09-08},
  year         = {2025}
}

@inproceedings{chen2018tvm,
  title     = {TVM: An automated end-to-end optimizing compiler for deep learning},
  author    = {Tianqi Chen and Thierry Moreau and Ziheng Jiang and Lianmin Zheng and Eddie Yan and Haichen Shen and Meghan Cowan and Leyuan Wang and Yuwei Hu and Luis Ceze and Carlos Guestrin and Arvind Krishnamurthy},
  booktitle = {Proceedings of the 13th USENIX Symposium on Operating Systems Design and Implementation (OSDI)},
  year      = {2018},
  pages     = {578--594}
}

@article{lattner2020mlir,
  title     = {MLIR: A compiler infrastructure for the end of Moore’s Law},
  author    = {Chris Lattner and others},
  journal   = {arXiv preprint arXiv:2002.11054},
  year      = {2020}
}

@article{rotem2018glow,
  title     = {Glow: Graph lowering compiler techniques for neural networks},
  author    = {Nadav Rotem and others},
  journal   = {arXiv preprint arXiv:1805.00907},
  year      = {2018}
}

@inproceedings{jia2019taso,
  title     = {TASO: Optimizing deep learning computation with automatic generation of graph substitutions},
  author    = {Zhihao Jia and Sina Lin and Charles R. Qi and Alex Aiken},
  booktitle = {Proceedings of the 27th ACM Symposium on Operating Systems Principles (SOSP)},
  pages     = {47--62},
  year      = {2019}
}

@misc{openxla2023,
  title     = {StableHLO and OpenXLA},
  author    = {OpenXLA Community},
  year      = {2023},
  url       = {https://openxla.org}
}

@misc{onnx2018,
  title     = {ONNX Runtime: High performance inference engine},
  author    = {Microsoft},
  year      = {2018},
  url       = {https://onnxruntime.ai}
}

@misc{agrawal2019tensorfloweagermultistagepythonembedded,
      title={TensorFlow Eager: A Multi-Stage, Python-Embedded DSL for Machine Learning}, 
      author={Akshay Agrawal and Akshay Naresh Modi and Alexandre Passos and Allen Lavoie and Ashish Agarwal and Asim Shankar and Igor Ganichev and Josh Levenberg and Mingsheng Hong and Rajat Monga and Shanqing Cai},
      year={2019},
      eprint={1903.01855},
      archivePrefix={arXiv},
      primaryClass={cs.PL},
      url={https://arxiv.org/abs/1903.01855}, 
}

@misc{jaseci2025,
  title        = {Jaseci: The Official Jaseci Code Repository},
  author       = {{Jaseci Labs}},
  year         = {2025},
  howpublished = {\url{https://github.com/Jaseci-Labs/jaseci}},
}

@misc{huggingface2025,
  author       = {{Hugging Face}},
  title        = {Hugging Face: Open-Source AI Community and Tools},
  year         = {2025},
  howpublished = {\url{https://huggingface.co}},
  note         = {Accessed: 2025-09-12}
}

@misc{pytorch_profiler,
  title        = {PyTorch Profiler: A performance debugging and analysis tool for PyTorch},
  howpublished = {\url{https://pytorch.org/docs/stable/profiler.html}},
  year         = {2021},
  note         = {accessed: 2025-09-12}
}

@article{biogpt,
    author = {Luo, Renqian and Sun, Liai and Xia, Yingce and Qin, Tao and Zhang, Sheng and Poon, Hoifung and Liu, Tie-Yan},
    title = "{BioGPT: generative pre-trained transformer for biomedical text generation and mining}",
    journal = {Briefings in Bioinformatics},
    volume = {23},
    number = {6},
    year = {2022},
    month = {09},
    issn = {1477-4054},
    doi = {10.1093/bib/bbac409},
    url = {https://doi.org/10.1093/bib/bbac409},
    note = {bbac409},
    eprint = {https://academic.oup.com/bib/article-pdf/23/6/bbac409/47144271/bbac409.pdf},
}

@misc{facebook_blenderbot_400m_distill,
  title        = {facebook/blenderbot-400M-distill},
  author       = {{Meta AI}},
  howpublished = {\url{https://huggingface.co/facebook/blenderbot-400M-distill}},
  year         = {2020}
}

@misc{flant5,
  doi = {10.48550/ARXIV.2210.11416},
  
  url = {https://arxiv.org/abs/2210.11416},
  
  author = {Chung, Hyung Won and Hou, Le and Longpre, Shayne and Zoph, Barret and Tay, Yi and Fedus, William and Li, Eric and Wang, Xuezhi and Dehghani, Mostafa and Brahma, Siddhartha and Webson, Albert and Gu, Shixiang Shane and Dai, Zhuyun and Suzgun, Mirac and Chen, Xinyun and Chowdhery, Aakanksha and Narang, Sharan and Mishra, Gaurav and Yu, Adams and Zhao, Vincent and Huang, Yanping and Dai, Andrew and Yu, Hongkun and Petrov, Slav and Chi, Ed H. and Dean, Jeff and Devlin, Jacob and Roberts, Adam and Zhou, Denny and Le, Quoc V. and Wei, Jason},
  
  title = {Scaling Instruction-Finetuned Language Models},
  
  publisher = {arXiv},
  
  year = {2022},
  
  copyright = {Creative Commons Attribution 4.0 International}
}

@article{Beltagy2020Longformer,
  title={Longformer: The Long-Document Transformer},
  author={Iz Beltagy and Matthew E. Peters and Arman Cohan},
  journal={arXiv:2004.05150},
  year={2020},
}

@misc{babybirdprd_moe_minicpm_x4_base,
  title        = {babybirdprd/moe-minicpm-x4-base},
  author       = {babybirdprd},
  howpublished = {\url{https://huggingface.co/babybirdprd/moe-minicpm-x4-base}},
  year         = {2025}
}

@misc{microsoft_phi4_mini_instruct,
  title        = {Microsoft Phi-4-mini-instruct},
  author       = {Microsoft},
  howpublished = {\url{https://huggingface.co/microsoft/Phi-4-mini-instruct}},
  year         = {2025}
}

@article{Qwen-Audio,
  title={Qwen-Audio: Advancing Universal Audio Understanding via Unified Large-Scale Audio-Language Models},
  author={Chu, Yunfei and Xu, Jin and Zhou, Xiaohuan and Yang, Qian and Zhang, Shiliang and Yan, Zhijie  and Zhou, Chang and Zhou, Jingren},
  journal={arXiv preprint arXiv:2311.07919},
  year={2023}
}

@misc{hf_tiny_random_pegasus_for_causallm,
  title        = {hf-internal-testing/tiny-random-PegasusForCausalLM},
  author       = {Hugging Face},
  howpublished = {\url{https://huggingface.co/hf-internal-testing/tiny-random-PegasusForCausalLM}},
  year         = {2025},
  note         = {Accessed: 2025-09-12; internal testing minimal model}
}

@inproceedings{autograph,title	= {AutoGraph: Imperative-style Coding with Graph-based Performance},author	= {Dan Moldovan and James Decker and Fei Wang and Andrew Johnson and Brian Lee and Zack Nado and D Sculley and Tiark Rompf and Alexander B Wiltschko},year	= {2019},booktitle	= {SysML}}

@inproceedings {MagPy,
author = {Chen Zhang and Rongchao Dong and Haojie Wang and Runxin Zhong and Jike Chen and Jidong Zhai},
title = {{MAGPY}: Compiling Eager Mode {DNN} Programs by Monitoring Execution States},
booktitle = {2024 USENIX Annual Technical Conference (USENIX ATC 24)},
year = {2024},
isbn = {978-1-939133-41-0},
address = {Santa Clara, CA},
pages = {683--698},
url = {https://www.usenix.org/conference/atc24/presentation/zhang-chen},
publisher = {USENIX Association},
month = jul
}

@inproceedings{TensoSSA,
author = {Ma, Jinming and Li, Xiuhong and Wang, Zihan and Zhang, Xingcheng and Yan, Shengen and Chen, Yuting and Zhang, Yueqian and Jin, Minxi and Jiang, Lijuan and Liang, Yun (Eric) and Yang, Chao and Lin, Dahua},
title = {A Holistic Functionalization Approach to Optimizing Imperative Tensor Programs in Deep Learning},
year = {2024},
isbn = {9798400706011},
publisher = {Association for Computing Machinery},
address = {New York, NY, USA},
url = {https://doi.org/10.1145/3649329.3658483},
doi = {10.1145/3649329.3658483},
booktitle = {Proceedings of the 61st ACM/IEEE Design Automation Conference},
articleno = {200},
numpages = {6},
location = {San Francisco, CA, USA},
series = {DAC '24}
}

@misc{gray2019cudagraphs,
  author       = {Gray, Alan},
  title        = {Getting Started with {CUDA} Graphs},
  howpublished = {NVIDIA Technical Blog},
  year         = {2019},
  month        = sep,
  note         = {Accessed: 2026-04-06},
  url          = {https://developer.nvidia.com/blog/cuda-graphs/}
}

@misc{radford2022whisper,
  doi = {10.48550/ARXIV.2212.04356},
  url = {https://arxiv.org/abs/2212.04356},
  author = {Radford, Alec and Kim, Jong Wook and Xu, Tao and Brockman, Greg and McLeavey, Christine and Sutskever, Ilya},
  title = {Robust Speech Recognition via Large-Scale Weak Supervision},
  publisher = {arXiv},
  year = {2022},
  copyright = {arXiv.org perpetual, non-exclusive license}
}

@article{bart-base,
  author    = {Mike Lewis and
               Yinhan Liu and
               Naman Goyal and
               Marjan Ghazvininejad and
               Abdelrahman Mohamed and
               Omer Levy and
               Veselin Stoyanov and
               Luke Zettlemoyer},
  title     = {{BART:} Denoising Sequence-to-Sequence Pre-training for Natural Language
               Generation, Translation, and Comprehension},
  journal   = {CoRR},
  volume    = {abs/1910.13461},
  year      = {2019},
  url       = {http://arxiv.org/abs/1910.13461},
  eprinttype = {arXiv},
  eprint    = {1910.13461},
  bibsource = {dblp computer science bibliography, https://dblp.org}
}

@article{ansari2024chronos,
    title={Chronos: Learning the Language of Time Series},
    author={Ansari, Abdul Fatir and Stella, Lorenzo and Turkmen, Caner and Zhang, Xiyuan and Mercado, Pedro and Shen, Huibin and Shchur, Oleksandr and Rangapuram, Syama Syndar and Pineda Arango, Sebastian and Kapoor, Shubham and Zschiegner, Jasper and Maddix, Danielle C. and Mahoney, Michael W. and Torkkola, Kari and Gordon Wilson, Andrew and Bohlke-Schneider, Michael and Wang, Yuyang},
    journal={Transactions on Machine Learning Research},
    issn={2835-8856},
    year={2024},
    url={https://openreview.net/forum?id=gerNCVqqtR}
}

@article{xiao2023florence,
  title={Florence-2: Advancing a unified representation for a variety of vision tasks},
  author={Xiao, Bin and Wu, Haiping and Xu, Weijian and Dai, Xiyang and Hu, Houdong and Lu, Yumao and Zeng, Michael and Liu, Ce and Yuan, Lu},
  journal={arXiv preprint arXiv:2311.06242},
  year={2023}
}

@article{2020t5,
  author  = {Colin Raffel and Noam Shazeer and Adam Roberts and Katherine Lee and Sharan Narang and Michael Matena and Yanqi Zhou and Wei Li and Peter J. Liu},
  title   = {Exploring the Limits of Transfer Learning with a Unified Text-to-Text Transformer},
  journal = {Journal of Machine Learning Research},
  year    = {2020},
  volume  = {21},
  number  = {140},
  pages   = {1-67},
  url     = {http://jmlr.org/papers/v21/20-074.html}
}

@inproceedings{huguet-cabot-navigli-2021-rebel-relation,
    title = "{REBEL}: Relation Extraction By End-to-end Language generation",
    author = "Huguet Cabot, Pere-Llu{\'\i}s  and
      Navigli, Roberto",
    booktitle = "Findings of the Association for Computational Linguistics: EMNLP 2021",
    month = nov,
    year = "2021",
    address = "Punta Cana, Dominican Republic",
    publisher = "Association for Computational Linguistics",
    url = "https://aclanthology.org/2021.findings-emnlp.204",
    pages = "2370--2381",
}

@inproceedings{huang2022layoutlmv3,
  author={Yupan Huang and Tengchao Lv and Lei Cui and Yutong Lu and Furu Wei},
  title={LayoutLMv3: Pre-training for Document AI with Unified Text and Image Masking},
  booktitle={Proceedings of the 30th ACM International Conference on Multimedia},
  year={2022}
}

@article{MOLformer,
  year = {2022},
  title = {{Large-scale chemical language representations capture molecular structure and   properties}},
  author = {Ross, Jerret and Belgodere, Brian and Chenthamarakshan, Vijil and Padhi, Inkit and   Mroueh, Youssef and Das, Payel},
  journal = {Nature Machine Intelligence},
  doi = {10.1038/s42256-022-00580-7},
  pages = {1256--1264},
  number = {12},
  volume = {4}
}

@inproceedings{tiedemann2020opus,
  title={{OPUS-MT} -- Building open translation services for the World},
  author={Tiedemann, J{\"o}rg and Thottingal, Santhosh},
  booktitle={Proceedings of the 22nd Annual Conference of the European Association for Machine Translation},
  year={2020},
  address={Lisboa, Portugal}
}

@article{inc-t5,
author = {Greco, Salvatore and La Quatra, Moreno and Cagliero, Luca and Cerquitelli, Tania},
title = {Towards AI-Assisted Inclusive Language Writing in Italian Formal Communications},
year = {2025},
issue_date = {August 2025},
publisher = {Association for Computing Machinery},
address = {New York, NY, USA},
volume = {16},
number = {4},
issn = {2157-6904},
url = {https://doi.org/10.1145/3729237},
doi = {10.1145/3729237},
journal = {ACM Trans. Intell. Syst. Technol.},
month = jun,
articleno = {79},
numpages = {24}
}

@misc{liu2023grounding,
      title={Grounding DINO: Marrying DINO with Grounded Pre-Training for Open-Set Object Detection}, 
      author={Shilong Liu and Zhaoyang Zeng and Tianhe Ren and Feng Li and Hao Zhang and Jie Yang and Chunyuan Li and Jianwei Yang and Hang Su and Jun Zhu and Lei Zhang},
      year={2023},
      eprint={2303.05499},
      archivePrefix={arXiv},
      primaryClass={cs.CV}
}

@misc{zhang2025jasperstelladistillationsota,
      title={Jasper and Stella: distillation of SOTA embedding models}, 
      author={Dun Zhang and Jiacheng Li and Ziyang Zeng and Fulong Wang},
      year={2025},
      eprint={2412.19048},
      archivePrefix={arXiv},
      primaryClass={cs.IR},
      url={https://arxiv.org/abs/2412.19048}, 
}

@misc{clap-model,
  doi = {10.48550/ARXIV.2211.06687},
  
  url = {https://arxiv.org/abs/2211.06687},
  
  author = {Wu, Yusong and Chen, Ke and Zhang, Tianyu and Hui, Yuchen and Nezhurina, Marianna and Berg-Kirkpatrick, Taylor and Dubnov, Shlomo},
  
  title = {Large-scale Contrastive Language-Audio Pretraining with Feature Fusion and Keyword-to-Caption Augmentation},
  
  publisher = {arXiv},
  
  year = {2022},
  
  copyright = {Creative Commons Attribution 4.0 International}
}

@misc{rand2025pytorch_compile,
  author={Chaim Rand},
  title={Maximizing {AI/ML} Model Performance with {PyTorch} Compilation},
  howpublished={\url{https://towardsdatascience.com/maximizing-ai-ml-model-performance-with-pytorch-compilation/}},
  year={2025},
  note={Accessed: 2026-04-07}
}

@misc{hoyath2025pytorch_dynamo,
  author={Hoyath},
  title={{ML4LM} --- {PyTorch} --- What Not to Do in {PyTorch} Models for Better Performance (dynamo)},
  howpublished={\url{https://hoyath.medium.com/ml4lm-pytorch-what-not-to-do-in-pytorch-models-for-better-performance-dynamo-2e5c675dbec2}},
  year={2025},
  note={Accessed: 2026-04-07}
}

@misc{pytorch_compile_tutorial,
  title={Introduction to torch.compile --- {PyTorch} Tutorials},
  author={{PyTorch Contributors}},
  howpublished={\url{https://docs.pytorch.org/tutorials/intermediate/torch_compile_tutorial.html}},
  year={2025},
  note={Accessed: 2026-04-07}
}

@misc{pytorch_compiler_faq,
  title={Frequently Asked Questions --- {PyTorch} Documentation},
  author={{PyTorch Contributors}},
  howpublished={\url{https://docs.pytorch.org/docs/stable/torch.compiler_faq.html}},
  year={2025},
  note={Accessed: 2026-04-07}
}

@book{aho2006compilers,
  author={Aho, Alfred V. and Lam, Monica S. and Sethi, Ravi and Ullman, Jeffrey D.},
  title={Compilers: Principles, Techniques, and Tools},
  edition={2nd},
  year={2006},
  publisher={Addison-Wesley}
}

@misc{sglang_server_args,
  author       = {{SGLang Project}},
  title        = {Server Arguments},
  howpublished = {SGLang Documentation},
  year         = {2026},
  note         = {Accessed: 2026-06-09},
  url          = {https://sgl-project.github.io/advanced_features/server_arguments.html}
}

@misc{vllm_piecewise_compile,
  author       = {{vLLM Project}},
  title        = {{torch.compile} Integration},
  howpublished = {vLLM Documentation},
  year         = {2026},
  note         = {Accessed: 2026-06-09},
  url          = {https://docs.vllm.ai/en/latest/design/torch_compile/}
}

@misc{vllm-debug-compile,
  author       = {{vLLM Project}},
  title        = {How to Debug the {vLLM}--torch.compile Integration},
  howpublished = {vLLM Documentation},
  year         = {2026},
  url          = {https://docs.vllm.ai/en/stable/design/debug_vllm_compile/},
  note         = {Accessed: 2026-06-11}
}

@misc{sglang-pcg,
  author       = {{SGLang Project}},
  title        = {Piecewise {CUDA} Graph},
  howpublished = {SGLang Documentation},
  year         = {2026},
  url          = {https://docs.sglang.io/advanced_features/piecewise_cuda_graph.html},
  note         = {Accessed: 2026-06-11}
}

@misc{pytorch-fullgraph,
  author       = {{PyTorch Contributors}},
  title        = {Use \texttt{fullgraph=True} to Identify and Eliminate Graph Breaks --- {PyTorch} Documentation},
  howpublished = {PyTorch Documentation},
  year         = {2025},
  url          = {https://docs.pytorch.org/docs/stable/compile/programming_model.fullgraph_true.html},
  note         = {Accessed: 2026-06-11}
}

@misc{vllm-transformers-backend,
  author       = {{vLLM Project}},
  title        = {Transformers Backend Integration in {vLLM}},
  howpublished = {vLLM Documentation},
  year         = {2026},
  url          = {https://docs.vllm.ai/en/latest/models/supported_models/},
  note         = {Accessed: 2026-06-11}
}

@misc{pytorch_common_graph_breaks,
  author       = {{PyTorch Contributors}},
  title        = {Common Graph Breaks --- PyTorch Documentation},
  year         = {2025},
  howpublished = {\url{https://docs.pytorch.org/docs/stable/compile/programming_model.common_graph_breaks.html}},
  note         = {Accessed: 2026-06-11}
}

@misc{pytorch_issue_132635,
  author       = {{PyTorch Contributors}},
  title        = {{[Dynamo]} support logging via reorderable\_logging\_functions. Issue \#132635, pytorch/pytorch},
  year         = {2024},
  howpublished = {\url{https://github.com/pytorch/pytorch/issues/132635}},
  note         = {Accessed: 2026-06-11}
}

@misc{torchcond,
  author = {{PyTorch Contributors}},
  title  = {Control Flow - Cond --- PyTorch Documentation},
  year   = {2026},
  howpublished = {\url{https://docs.pytorch.org/docs/2.12/higher_order_ops/cond.html}},
  note   = {Accessed: 2026-08-10}
}

@misc{torchcondimpl,
  author = {{PyTorch Contributors}},
  title  = {torch/\_higher\_order\_ops/cond.py, pytorch/pytorch},
  year   = {2026},
  howpublished = {\url{https://github.com/pytorch/pytorch/blob/main/torch/_higher_order_ops/cond.py}},
  note   = {Accessed: 2026-08-10}
}

\end{document}